\documentclass{sig-alternate-05-2015}
\usepackage{graphicx}
\usepackage{subfig}
\usepackage{makecell}
\usepackage{tabularx}
\usepackage{balance}
\usepackage{ragged2e}
\usepackage{xcolor}
\newcommand{\name}{{\sc Blockbench}}
\begin{document}
\makeatletter
\def\@copyrightspace{\relax}
\makeatother

\title{BLOCKBENCH: A Framework for Analyzing Private Blockchains}
\numberofauthors{1}
\author{
  \alignauthor  \mbox{Tien Tuan Anh Dinh$^{\ddag}$ \hspace{0.1em} Ji Wang$^{\ddag}$ \hspace{0.1em} Gang
  Chen$^{\S}$ \hspace{0.1em} Rui Liu$^{\ddag}$ \hspace{0.1em} Beng Chin Ooi$^{\ddag}$ \hspace{0.1em} Kian-Lee Tan$^{\ddag}$}\\
  \affaddr{$^\ddag$ National University of Singapore \qquad $^\S$ Zhejiang University}\\
  \email{$^{\ddag}$ \{dinhtta, wangji, liur, ooibc, tankl\}@comp.nus.edu.sg \qquad $^\S$ cg@zju.edu.cn}
}

\maketitle

\begin{abstract}
Blockchain technologies are taking the world by storm. Public blockchains, such as Bitcoin and Ethereum, enable secure
peer-to-peer applications like crypto-currency or smart contracts. Their security and performance are well studied.
This paper concerns recent private blockchain systems designed with stronger security (trust) assumption and
performance requirement. These systems target and aim to disrupt applications which have so far been implemented on
top of database systems, for example banking, finance and trading applications. Multiple platforms for private
blockchains are being actively developed and fine tuned. However, there is a clear lack of a systematic framework with
which different systems can be analyzed and compared against each other. Such a framework can be used to assess
blockchains' viability as another distributed data processing platform, while helping developers to identify
bottlenecks and accordingly improve their platforms.

In this paper, we first describe \name, the first evaluation framework for analyzing private blockchains. It serves as a
fair means of comparison for different platforms and enables deeper understanding of different system design choices.
Any private blockchain can be integrated to \name\ via simple APIs and benchmarked against workloads that are based on
real and synthetic smart contracts. \name\ measures overall and component-wise performance in terms of throughput,
latency, scalability and fault-tolerance. Next, we use \name\ to conduct comprehensive evaluation of three major private
blockchains: Ethereum, Parity and Hyperledger Fabric. The results demonstrate that these systems are still far from
displacing current database systems in traditional data processing workloads. Furthermore, there are gaps in performance
among the three systems which are attributed to the design choices at different layers of the blockchain's software
stack.  \end{abstract}

\section{Introduction}
\label{sect:intro}
Blockchain technologies are gaining massive momentum in the last few years, largely due to the success of
Bitcoin crypto-currency~\cite{nakamoto2008bitcoin}. A blockchain, also called distributed ledger, is essentially an
append-only data structure maintained by a set of nodes which do not fully trust each other. All nodes in a blockchain
network agree on an ordered set of blocks, each containing multiple transactions, thus the blockchain can be viewed as a
log of ordered transactions. In a database context, blockchain can be viewed as a solution to the distributed
transaction management problems: nodes keep replicas of the data and agree on an execution order of transactions.
However, traditional database systems work in a trusted environment and employ well known concurrency control
techniques~\cite{qian16,thomson12,bailis14} to order transactions.  Blockchain's key advantage is that it does not
assume nodes trust each other and therefore is designed to achieve Byzantine fault tolerance.    
 
In the original design, Bitcoin's blockchain stores {\em coins} as the system states shared by all participants. For
this simple application, Bitcoin nodes implement a simple replicated state machine model which simply moves coins from
one address to another. Since then, blockchain has grown rapidly to support user-defined states and Turing complete
state machine models. Ethereum~\cite{ethereum} is a well-known example which enables any decentralized, replicated
applications known as {\em smart contracts}. More importantly, interest from the industry has started to drive
development of new blockchain platforms that are designed for private settings in which participants are authenticated.
Blockchain systems in such environments are called private (or {\em permissioned}), as opposed to the early systems
operating in public (or {\em permissionless}) environments where anyone can join and leave. Applications for security
trading and settlement~\cite{ripple}, asset and finance management~\cite{melonport,morgan16}, banking and insurance~\cite{gs16}
are being built and evaluated. These applications are currently supported by enterprise-grade database systems like
Oracle and MySQL, but blockchain has the potential to disrupt this status quo because it incurs lower infrastructure and
human costs~\cite{gs16}. { In particular, blockchain's immutability and transparency help reduce human
errors and the need for manual intervention due to conflicting data. Blockchain can help streamline business processes
by removing duplicate efforts in data governance.} Goldman Sachs estimated 6 billion saving in current capital
market~\cite{gs16}, and J.P.  Morgan forecast that blockchains will start to replace currently redundant infrastructure
by 2020~\cite{morgan16}.     

Given this trend in employing blockchain in settings where database technologies have established dominance, one
question to ask is to what extent can blockchain handle data processing workload. Another question is which platform to
choose from many that are available today, because even though blockchain is an open protocol, different platforms
exist in silo.  In this work, we develop a benchmarking framework called \name\ to address both questions. \name\ is the
first benchmark for studying and comparing performance of permissioned blockchains. Although nodes in a
permissioned blockchain still do not trust each other, their identities are authenticated, which allows the
system to use more efficient protocols for tolerating Byzantine failure than in public settings.  We do not
focus on public blockchains because their performance (and trade-offs against security guarantee) is
relatively well studied~\cite{gervais16, Luu2015demystifying, croman2016scaling, bonneau2015sok}. Our
framework is not only useful for application developers to assess blockchain's potentials in meeting the
application need, but also offers insights for platform developers: helping them to identify and improve on
the performance bottlenecks. 

We face three challenges in developing \name. First, a blockchain system comprises many parts and we observe that a wide
variety of design choices are made among different platforms at almost every single detail. In \name, we divide
the blockchain architecture into three modular layers and focus our study on them: the consensus layer, data model and
execution layer. Second, there are many different choices of platforms, but not all of them have reached a mature design,
implementation and an established user base. For this, we start by designing \name\ based on three most mature platforms
within our consideration, namely Ethereum~\cite{ethereum}, Parity~\cite{parity} and Hyperledger~\cite{hyperledger}, and
then generalize to support future platforms. 
All three platforms support smart contracts and can be deployed in a private environment. Third, there is lack of a
database-oriented workloads for blockchain. 
Although the real Ethereum transactions and contracts can be found on the
public blockchain, it is unclear if such workload is sufficiently representative to assess blockchain's general data processing
capabilities.  To address this challenge, we treat blockchain as a key-value storage coupled with an engine which can
realize both transactional and analytical functionality via smart contracts. We then design and run both transaction and
analytics workloads based on real and synthetic data.

\name\ is a flexible and extensible framework that provides a number of workloads, and comes with Ethereum, Parity and
Hyperledger as backends. Workloads are transaction-oriented currently and designed to macro-benchmark and
micro-benchmark blockchain for supporting database-like applications. Specifically, the current macro-benchmark includes
a key-value (YCSB), an OLTP (Smallbank) workload and a number of real Ethereum smart contract workloads. For each of the
consensus, data and execution layer, there is at least a micro-benchmark workload to measure its performance in
isolation. For example, for the execution layer, \name\ provides two workloads that stress test the smart contract I/O
and computation speed.  New workloads and blockchains can be easily integrated via a simple set of APIs. \name\
quantifies the performance of a backend system in several dimensions: throughput, latency, scalability and fault
tolerance. It supports security evaluation by simulating network-level attacks. Using \name, we conduct an in-depth
comparison of the three blockchain systems on two macro benchmark and four micro benchmark workloads. The results show
that blockchain systems' performance is limited, far below what is expected of a state-of-the-art database system (such
as H-Store).  Hyperledger consistently outperforms the other two systems across seven benchmarks. But it fails to scale
beyond 16 nodes. Our evaluation shows that the consensus protocols account for the performance gap at the application
layer for Ethereum and Hyperledger. We also identify a processing bottleneck in Parity. Finally, our evaluation also
reveals bottlenecks in the execution and data layer of Ethereum and Parity. 

In summary, our contributions are:
\begin{itemize}
\item We present the first benchmark for understanding and comparing the performance of permissioned blockchain systems. 
\item We conduct a comprehensive evaluation of Ethereum, Parity and Hyperledger. Our empirical results present concrete
evidence of blockchain's limitations in handling data processing workloads, and reveal bottlenecks in the three systems.
The results serve as a baseline for further development of blockchain technologies.  \end{itemize}

In the next section, we discuss blockchain systems in more detail. Section~\ref{sec:design} describes \name\ design and
implementation. Section~\ref{sec:evaluation} presents our comparative performance studies of three systems. We discuss
lessons learned from the results in Section~\ref{sec:discussion} and related work in Section~\ref{sec:related},
and we conclude in Section~\ref{sec:conclusion}.    

\section{Private BlockChains}
\label{sec:background}

A typical blockchain system consists of multiple nodes which do not fully trust each other. Some nodes exhibit Byzantine
behavior, but the majority is honest. Together, the nodes maintain a set of shared, global states and perform
transactions modifying the states. Blockchain is a special data structure which maintains the states and the historical
transactions.  All nodes in the system agree on the transactions and their order as stored on the blockchain. Because of
this, blockchain is often referred to as a distributed ledger.  

{\bf Blockchain transactions.} A transaction in a blockchain is the same as in traditional database: a sequence of
operations applied on some states. As such, a blockchain transaction requires the same ACID semantics. The key
difference is the failure model under consideration. Current transactional, distributed databases~\cite{hstore,
spanner} employ classic concurrency control techniques such as two-phase commit to ensure ACID.  They can achieve high
performance, because of the simple failure model, i.e. crash failure.  In contrast, the original blockchain design considers
a much hostile environment in which nodes are Byzantine and they are free to join and leave. Under this
model, the overhead of concurrency control is much higher~\cite{castro1999practical}.       

\begin{figure}
\centering
\includegraphics[scale=0.4]{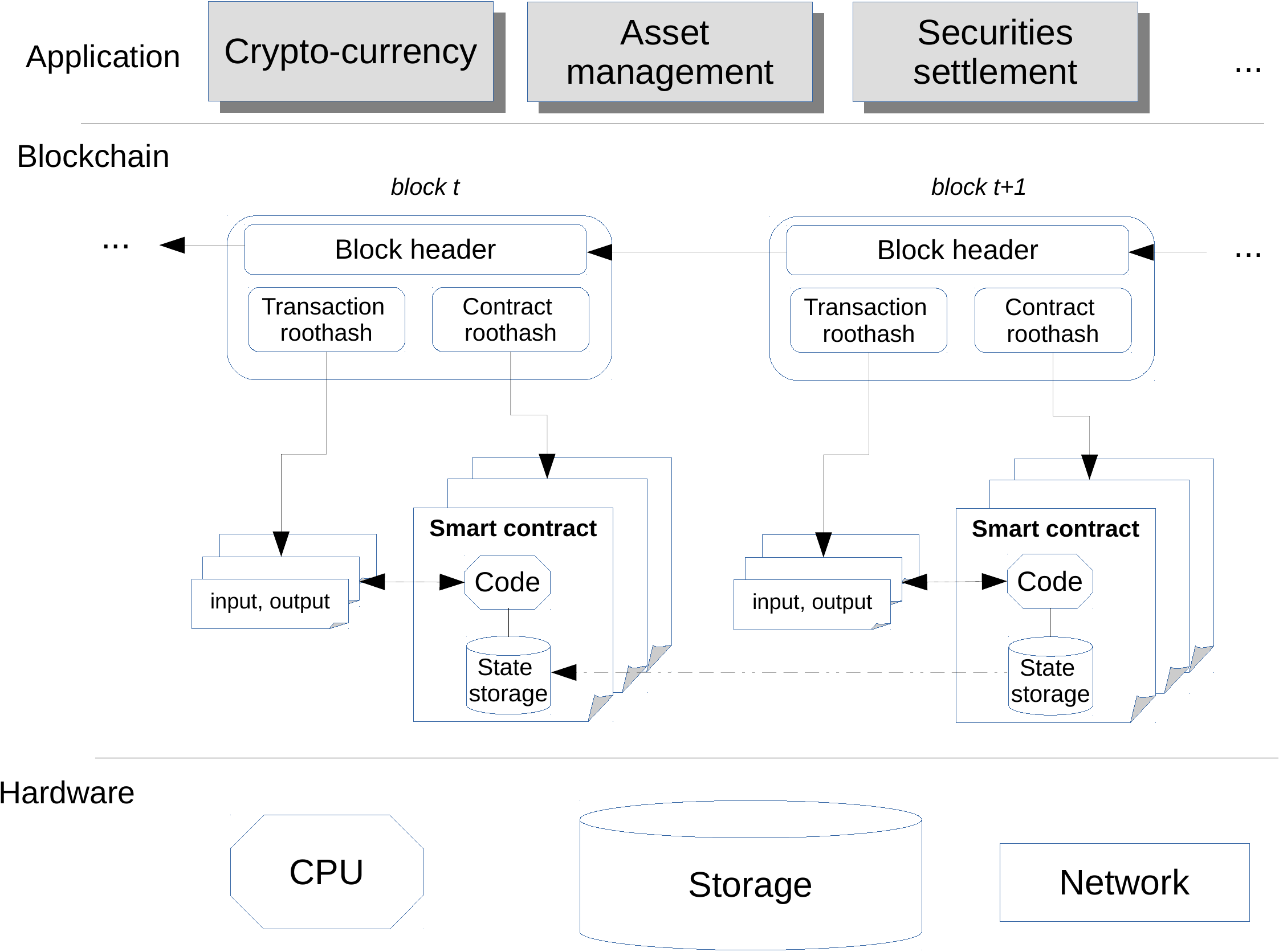}
\caption{Blockchain software stack on a fully validating node. A non-validating node stores only the block headers.
Different blockchain platforms offer different interface between the blockchain and application layer.}
\label{fig:stack}
\end{figure}

{\bf Bitcoin.} In Bitcoin~\cite{nakamoto2008bitcoin}, the states are digital coins (crypto-currency) available in the network. A
Bitcoin transaction moves coins from one set of addresses to another set of addresses. Each node broadcasts a set of
transactions it wants to perform.  Special nodes called {\em miners} collect transactions into blocks, check for
their validity, and start a consensus protocol to append the blocks onto the blockchain. Figure~\ref{fig:stack}
shows the blockchain data structure, in which each block is linked to its predecessor via a cryptographic pointer, all
the way back to the first (genesis) block. Bitcoin uses {\em proof-of-work} (PoW) for consensus: only a miner
which has successfully solved a computationally hard puzzle (finding the right nonce for the block header) can append to the
blockchain. PoW is tolerant of Byzantine failure, but it is probabilistic in nature: it is possible that two blocks are appended
at the same time, creating a {\em fork} in the blockchain.  
Bitcoin resolves this by only considering a block as confirmed
after it is followed by a number of blocks (typically six blocks). This probabilistic guarantee causes both security and
performance issues: attacks have been demonstrated by an adversary controlling only 25\% of the nodes~\cite{eyal14}, and
Bitcoin transaction throughput remains very low (7 transactions per second~\cite{croman2016scaling}).  

\begin{figure}
\centering
{\footnotesize
\begin{verbatim}
contract Doubler{
  struct Partitipant {
    address etherAddress;
    uint amount; 
  }
  Partitipant[] public participants;
  unit public balance = 0; 
  ...
  function enter(){
    ...
    balance+= msg.value; 
    ...
    if (balance > 2*participants[payoutIdx].amount){
      transactionAmount = ...
      participants[payoutIdx].
        etherAddress.send(transactionAmount); 
      ...
    }
  }
  ...
}
\end{verbatim}
}
\caption{An example of smart contract, written in Solidity language, for a pyramid scheme on Ethereum.}
\label{fig:doubler}
\end{figure}
{\bf Ethereum.} Due to simple transaction semantics, Bitcoin nodes execute a very simple state machine pre-built into
the protocol.  Ethereum~\cite{ethereum} extends Bitcoin to support user-defined and Turing complete state machines.
In particular, Ethereum blockchain lets the user define any complex computations in the form of smart contracts. Once
deployed, the smart contract is executed on all Ethereum nodes as a replicated state machine. Beside the shared states of
the blockchains (crypto-currency, for example), each smart contract has access to its own states. Figure~\ref{fig:stack}
shows the software stack in a typical Ethereum node: a fully validating node contains the entire history of the
blockchain, whereas a non-validating node stores only the block headers.  One key difference with Bitcoin is that smart
contract states are maintained as well as normal transactions. In fact, a smart contract is identified by a unique
address which has its own money balance (in Ether), and upon retrieving a transaction to its address, it executes the
contract's logics. Ethereum comes with an execution engine, called Ethereum Virtual Machine (EVM), to execute smart
contracts.  Figure~\ref{fig:doubler} shows a snippet of popular contract running on Ethereum, which implements a pyramid
scheme: users send money to this contract which is used to pay interests to early participants. This contract has its
own states, namely the list of participants, and exports a function called {\tt enter}. A user invokes this contract by
sending his money through a transaction, which is accessed by the smart contract as {\tt msg.sender} and {\tt
msg.amount}. 

{\bf Private blockchain.} Ethereum uses the same consensus protocol as Bitcoin does, though with different parameters.
In fact, 90\% of public blockchain systems employ variants of the proof-of-work protocol.  PoW is non-deterministic and
computationally expensive, both rendering it unsuitable for applications such as banking and finance which must handle a
lot of transactions in a deterministic manner. Recent blockchain systems, e.g., Hyperledger, consider restricted settings wherein nodes are
authenticated. Although PoW is still useful in such permissioned environments, as in the case of Ethereum, there are more
efficient and deterministic solutions where node identities are known. Distributed fault-tolerant consensus in such a
closed settings is a well studied topic in distributed systems. Zab~\cite{zab}, Raft~\cite{raft}, Paxos~\cite{paxos},
PBFT~\cite{castro1999practical} are popular protocols that are in active use today. Recent permissioned blockchains
either use existing PBFT, as in Hyperledger~\cite{hyperledger}, or develop their own variants, as in
Parity~\cite{parity}, Ripple~\cite{ripple} and ErisDB~\cite{monax}. Most of these systems support smart contracts,
though in different languages, with different APIs and execution engines (see a more comprehensive comparison in the
Appendix). As a result, permissioned blockchains can execute complex application more efficiently than PoW-based
blockchains, while being Byzantine fault tolerant. These properties and the commercial interests from major banking and
financial institutions have bestowed on private blockchains the potentials to disrupt the current practice in data
management.

\section{\name\ Design}
\label{sec:design}
This section discusses blockchain's common layers of abstractions and the benchmarking workloads. 

\begin{figure}
\centering
\includegraphics[scale=0.36]{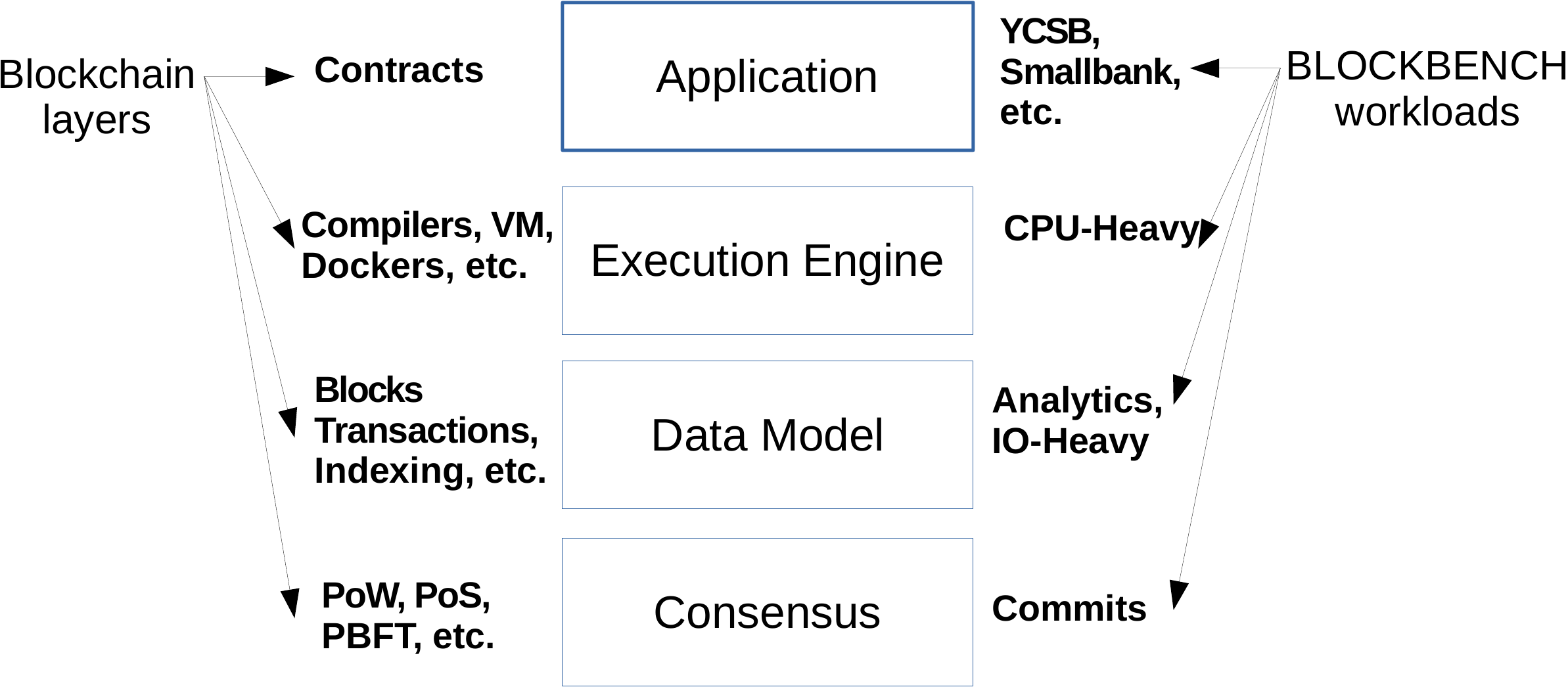}
\caption{Abstraction layers in blockchain, and the corresponding workloads in \name.}
\label{fig:layer}
\end{figure}

\newpage

\subsection{Blockchain Layers}
There are many choices of blockchains: over 200 Bitcoin variants, Ethereum and other permissioned blockchains.  To
meaningfully compare them, we identify four abstraction layers found in all of these systems (Figure~\ref{fig:layer})
and design our workloads to target these layers. The consensus layer contains protocols via which a block is considered
appended to the blockchain. The data layer contains the structure, content and operations on the blockchain data. The
execution layer includes details of the runtime environment support blockchain operations. Finally, the application
layer includes classes of blockchain applications. In a related work, Croman et. al.~\cite{croman2016scaling}
proposed to divide blockchain into several planes: network, consensus, storage, view and side plane. While
similar to our four layers, the plane abstractions were geared towards crypto-currency applications and did
not take into account the execution of smart contracts. Our layers model more accurately the real
implementations of private blockchains. We now discuss these layers in turn. 

\subsubsection{Consensus} The role of the consensus layer is to get all nodes in the system to agree on the blockchain
content. That is, if a node appends (or commits) a block, the other nodes also append the same block to their copy of
the blockchain.  Protocols for reaching consensus in the crash-failure model play a key role in distributed databases,
wherein nodes agree on a global transaction order. Blockchain systems, on the other hand, employ a spectrum of Byzantine
fault-tolerant protocols~\cite{vulkolic15}.  

At one extreme, Ethereum, like Bitcoin, uses proof-of-work whose difficulty is agreed upon and adjusted gradually to
achieve a rate of (currently) one block per $14s$ (Bitcoin's difficulty achieves a rate of one block per $10m$).
In essence, proof-of-work selects at each round a random node which can append a block, where the probability of being
selected is determined by the node's total computing power. {This simple scheme works against Sybil attack~\cite{sybil} -
a common attack in open, decentralized environments in which the adversary can acquire multiple identities.} However, it
consumes a lot of energy and computing power, as nodes spend their CPU cycles solving puzzles instead of doing otherwise
useful works. Worse still, it does not guarantee safety: two nodes may both be selected to append to the blockchain, and
both blocks can be accepted. This causes fork in the blockchain, and most PoW-based systems add additional rules, for
example, only blocks on the longest chain are considered accepted.  Ethereum, in particular,  adopts a PoW variant
called GHOST~\cite{ghost} which accepts blocks in heavy branches. In any case, a block can be confirmed as part of the
blockchain only with some high probability. 

At the other extreme, Hyperledger uses the classic PBFT protocol, which is communication bound: $O(N^2)$ where N is the
number of nodes. PBFT can tolerate fewer than $\frac{N}{3}$ failures, and works in three phases in which nodes broadcast
messages to each other. First, the {\em pre-prepare} phase selects a leader which chooses a value to commit. Next, the
{\em prepare} phase broadcasts the value to be validated. Finally, the {\em commit} phase waits for more than two third of
the nodes to confirm before announcing that the value is committed. PBFT has been shown to achieve liveness and safety properties
in a partially asynchronous model \cite{castro1999practical}, thus, unlike PoW, once the block is appended it is confirmed immediately. It can
tolerate more failures than PoW (which is shown to be vulnerable to $25\%$ attacks~\cite{eyal14}). However, PBFT
assumes that node identities are known, therefore it can only work in the permissioned settings. Additionally, the protocol
is unlikely to be able to scale to the network size of Ethereum, because of its communication overhead.  

In between, there are various hybrid designs that combine both scalability of PoW and safety property of
PBFT~\cite{hybridconsensus}. For example, Bitcoin-NG~\cite{bitcoin-ng} decouples consensus from transaction validation
by using PoW for leader election who can then append more than one block at a time. Similarly, Byzcoin~\cite{byzcoin} and
Elastico~\cite{elastico} leverage PoW to determine random, smaller consensus groups which run PBFT. Another example is
the Tendermint protocol, adopted by ErisDB~\cite{monax}, which combines proof-of-stake (PoS) and PBFT. Unlike PoW, PoS
selects a node which can append a block by its investment (or stake) in the system, therefore avoid expending CPU
resources. Parity~\cite{parity} implements a simplified version of PoS called Proof of Authority (or PoA). In this
protocol, a set of {\em authorities} are pre-determined and each authority is assigned a fixed time slot within which it
can generate blocks.  PoA makes a strong assumption that the authorities are trusted, and therefore is only suitable for
private deployment. 

\subsubsection{Data model} In Bitcoin, transactions are first class citizens: they are system states representing
digital coins in the network. Private blockchains depart from this model, by focusing on {\em accounts}. One immediate
benefit is simplicity, especially for applications involving crypto-currencies. For instance, transferring money from
one user to another in Bitcoin involves searching for transactions belonging to the sender, then marking some of them as
spent, whereas it is easily done in Ethereum by updating two accounts in one transaction. An account in Ethereum has a
balance as its state, and is updated upon receiving a transaction. A special type of account, called {\em smart
contract}, contains executable code and private states (Figure~\ref{fig:stack}). When receiving a transaction, in
addition to updating its balance, the contract's code is invoked with arguments specified in the transaction.  The code
can read the states of other non-contract accounts, and it can send new transactions during execution. Parity adopts the
same data model as in Ethereum.  In Hyperledger, there is only one type of account called {\em chaincode} which is the
same as Ethereum's contract. Chaincode can only access its private storage and they are isolated from each other.  

A block contains a list of transactions, and a list of smart contracts executed as well as their latest states.
Each block is identified by the cryptographic hash of its content, and linked to the previous block's identity. In
Parity, the entire block content is kept in memory. In Ethereum and Hyperledger, the content is organized in a two
layered data structure. The states are stored in a disk-based key-value storage (LevelDB\cite{leveldb} in Ethereum and
RocksDB\cite{rocksdb} in Hyperledger), and organized in a hash tree whose root is included in the block header. Ethereum
caches the states in memory, while Hyperledger outsources its data management entirely to the storage engine. Only
states affected by the block's transactions are recorded in the root hash. The hash tree for transaction list is a
classic Merkle tree, as the list is not large. On the other hand, different Merkle tree variants are used for the state
tree. Ethereum and Parity employ Patricia-Merkle tree that supports efficient update and search operations.  Hyperledger
implements Bucket-Merkle tree which uses a hash function to group states into a list of buckets from which a Merkle tree
is built.  

Block headers and the key-value storage together maintain all the historical transactions and states of the blockchain. For
validating and executing transactions, a blockchain node needs only a few recent blocks (or just the latest block for
PBFT-based systems). However, the node also interacts via some RPC-like mechanisms with light-weight clients who do not
have the entire blockchain. Such external interfaces enable building of third-party applications on top of blockchain.
Current systems support a minimum set of queries including getting blocks and transactions based on their IDs.  Ethereum
and Parity expose a more comprehensive set of APIs via JSON-RPC, supporting queries of account states at specific blocks
and of other block statistics.  

\subsubsection{Execution layer} A contract (or chaincode) is executed in a runtime environment. One requirement is that
the execution must be fast, because there are multiple contracts and transactions in one block and they must all be
verified by the node. Another is that the execution must be deterministic, ideally the same at all nodes. Deterministic
execution avoid unnecessary inconsistency in transaction input and output which leads to blocks being aborted. In both
PoW and PBFT, aborting transactions wastes computing resources. 

Ethereum develops its own machine language (bytecode) and a virtual machine (called EVM) for executing the code, which
is also adopted by Parity. EVM is optimized for Ethereum-specific operations. For example, every code instruction
executed in Ethereum costs a certain amount of {\em gas}, and the total cost must be properly tracked and charged to the
transaction's sender.  Furthermore, the code must keep track of intermediate states and reverse them if the execution
runs out of gas.  Hyperledger, in contrast, does not consider these semantics in its design, so it simply supports
running of compiled machine codes inside Docker images.  Specifically, chaincodes are deployed as Docker images
interacting with Hyperledger's backend via pre-defined interfaces. One advantage of Hyperledger's environment is that it
supports multiple high-level programming languages such as Go and Java, as opposed to Ethereum's own language. In terms
of development environment, Hyperledger exposes only simple key-value operations, namely {\tt putState} and {\tt
getState}. This is restricted, because any contract states must be mapped into key-value tuples. In contrast, Ethereum
and Parity support a richer set of data types  such as map, array and composite structures. These high-level data types
in Ethereum and Parity make it easier and faster to develop new contracts.  

\subsubsection{Application layer} Many applications are being proposed for blockchain, leveraging the latter's two key
properties.  First, data in the blockchain is immutable and transparent to the participants, meaning that once a record is
appended, it can never be changed.  Second, it is resilient to dishonest and malicious participants. Even in permissioned settings,
participants can be mutually distrustful. The most popular application, however, is still crypto-currency. Ethereum has
its own currency (Ether) and a majority of smart contracts running on it are currency related. Decentralized Autonomous
Organization (DAO) is the most active application in Ethereum, creating communities for crowd funding, exchange,
investment, or any other decentralized activities. A DAO manages funds contributed by participants and gives its users voting
power proportional to their contributions. Parity's main application is the wallet application that manages Ether. As
major banks are now considering adopting crypto-currency, some fintech companies are building applications that take
crypto-currency to mediate financial transactions, for example, in currency exchange market~\cite{ripple}. Other
examples include applying the currency and smart contracts for more transparent and cost-effective asset
management~\cite{melonport,morgan16}.  

Some applications propose to build on blockchain's immutability and transparency for better application workflows in
which humans are the bottlenecks. For example, security settlements and insurance processes can be sped up by storing data
on the blockchain~\cite{gs16}. Another example is sharing economy applications, such as AirBnB, which can use blockchain
to evaluate reputation and trust in a decentralized settings, because historical activities of any users are available
and immutable. This also extends to Internet of Things settings, where devices need to establish trust among each
other~\cite{watsoniot}.  

\subsection{\name\ Implementation}
\begin{figure}
\centering
\includegraphics[scale=0.4]{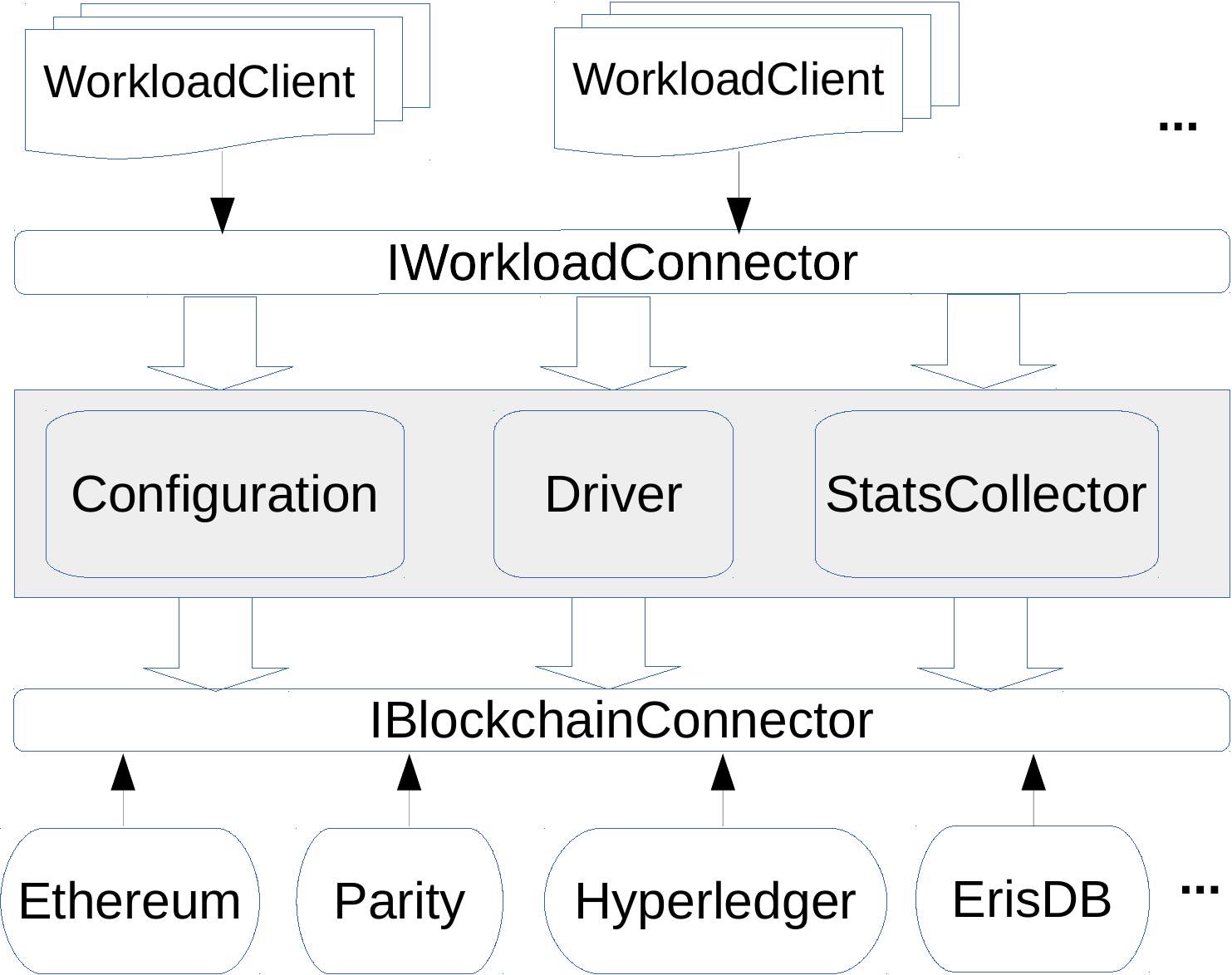}
\caption{\name\ software stack. New workloads are added by implementing {\tt IWorkloadConnector} interface. New
blockchain backends are added by implementing {\tt IBlockchainConnector}. Current backends include Ethereum, Parity and
Hyperledger.} 
\label{fig:blockbenchstack}
\end{figure}
Figure~\ref{fig:blockbenchstack} illustrates the current \name's implementation. To evaluate a blockchain system, the
first step is to integrate the blockchain into the framework's backend by implementing {\tt IBlockchainConnector} interface. The
interface contains operations for deploying application, invoking it by sending a transaction, and for querying
the blockchain's states. Ethereum, Parity and Hyperledger are current backends supported by \name, while ErisDB integration is under
development. A user can use one of the existing workloads (discussed next) to evaluate the blockchain, or implement a new
workload using the {\tt IWorkloadConnector} interface (we assume that the smart contract handling the workload's logic
is already implemented and deployed on the blockchain). This interface essentially wraps the workload's operations into
transactions to be sent to the blockchain. Specifically, it has a {\tt getNextTransaction} method which returns a new
blockchain transaction. \name's core component is the {\tt Driver} which takes as input a workload, user-defined
configuration (number of operations, number of clients, threads, etc.), executes it on the blockchain and outputs
running statistics. 

{\bf Asynchronous Driver.} One challenge in implementing the {\tt Driver} is that current blockchain systems are {\em
asynchronous services}, meaning that transactions submitted to the systems are processed at a later time. This is in
contrast to databases, especially transactional databases, in which operations are synchronous, i.e. they block until
the systems finish processing. When a transaction is submitted, Ethereum, Parity and Hyperledger return a transaction ID
which can be used for checking the transaction status at a later time. Such asynchronous semantics could result in
better performance, but it forces the {\tt Driver} to periodically poll for status of the submitted requests. In
particular, {\tt Driver} maintains a queue of outstanding transactions that have not been confirmed. New transaction IDs
are added to the queue by worker threads. A polling thread periodically invokes {\tt getLatestBlock(h)} method in the
{\tt IBlockchainConnector} interface, which returns a list of new {\em confirmed} blocks on the blockchain from a given
height $h$. Ethereum and Parity consider a block as confirmed if it is at least {\tt confirmationLength} blocks from the current
blockchain's tip, whereas Hyperledger confirms a block as soon as it appears on the blockchain.  The {\tt Driver} then
extracts transaction lists from the confirmed blocks' content and removes matching ones in the local queue. {\tt
getLatestBlock(h)} can be implemented in all three systems by first requesting for the blockchain's current tip $t$,
then requesting the content of all blocks in the range $(h,t]$. ErisDB provides a publish/subscribe interface that could
simplify the implementation of this function. 

\subsection{Evaluation Metrics}
\label{subsec:metrics}
The output statistics of running a workload with different configurations can be used to evaluate the blockchain against
three performance metrics.  

\begin{itemize}
\item Throughput: measured as the number of successful transactions per second. A workload can be configured
with multiple clients and threads per clients to saturate the blockchain throughput. 

\item Latency: measured as the response time per transaction. {\tt Driver} implements blocking transaction, i.e. it
waits for one transaction to finish before starting another. 

\item Scalability: measured as the changes in throughput and latency when increasing number of
nodes and number of concurrent workloads.  

\item Fault tolerance: measured as how the throughput and latency change during node failure. Although blockchain
systems are tolerant against Byzantine failure, it is not possible to simulate all Byzantine behaviors. In \name\ we
simulate three failure modes: crash failure in which a node simply stops, network delay in which we inject arbitrary
delays into messages, and random response in which we corrupt the messages exchanged among the nodes. 
\end{itemize}

{\textbf{Security metrics.} A special case of Byzantine failures that is important to blockchain systems is malicious
behavior caused by an attacker. The attacker can be a compromised node or rouge participant within the system. Under
this threat model, security of a blockchain is defined as the safety property of the underlying consensus protocol. In
particular, security means that the non-Byzantine nodes have the same blockchain data. Violation of the safety property
leads to forks in the blockchain. Classic Byzantine tolerant protocols such as PBFT are proven to ensure safety for a
certain number of failures, thus security is guaranteed. On the other hand, in PoW systems like Bitcoin or Ethereum,
forks can occur due to network delays causing two nodes to mine the same blocks. While such accidental forks can be
quickly resolved, forks engineered by the attackers can be used for {\em double spending} and {\em selfish mining}. In
the former, the attacker sends a transaction to a block in the fork, waits for it to be accepted by the users, then
sends a conflicting transaction to another block in the main branch. In the latter, by withholding blocks and
maintaining a private, long fork, the attacker disrupts the incentives for mining and forces other participants to join
the attacker's coalition. By compromising $25\%$ of the nodes, the attacker can control the entire network's block
generation~\cite{eyal14}. } 

{In this work we quantify security as the number of blocks in the forks. Such blocks, called orphan or stale blocks,
represent the window of vulnerability in which the attacker can perform double spending or selfish mining.  To
manipulate forks, the key strategy is to isolate a group of nodes, i.e. to partition the network. For example, eclipse
attack~\cite{eclipse} exploits the application-level protocol to surround the targeted nodes with ones under the
attacker's control. At the network level, BGP hijacking~\cite{bgp_hijack} requires controlling as few as $900$ prefixes
to isolate $50\%$ of the Bitcoin's total mining power. \name\ implements a simulation of these attacks by partitioning
the network for a given duration.  In particular, during partition \name\ runtime drops network traffic between any two
nodes in the two partitions. Security is then measured by the ratio between the total number of blocks included in the
main branch and the total number of blocks confirmed by the users. The lower the ratio, the less vulnerable the system
is from double spending for selfish mining.  }


\subsection{Workloads}
\label{subsec:workloads}
\begin{table}
\centering
\resizebox{\columnwidth}{!}{%
\begin{tabular}{|l|l|}
\hline
{\bf Smart contracts} & {\bf Description} \\ \hline
YCSB & Key-value store \\ \hline
Smallbank & OLTP workload \\ \hline
EtherId & Name registrar contract \\ \hline
Doubler & Ponzi scheme \\ \hline
WavesPresale & Crowd sale \\ \hline
VersionKVStore & Keep state's versions (Hyperledger only)\\ \hline
IOHeavy & Read and write a lot of data \\ \hline
CPUHeavy & Sort a large array \\ \hline
DoNothing & Simple contract, do nothing \\ \hline
\end{tabular}
}
\caption{Summary of smart contracts implemented in \name. Each contract has one Solidity version for Parity and
Ethereum, and one Golang version for Hyperledger.}
\label{tab:workloads}
\end{table}

We divide the workloads into two major categories: macro benchmark for evaluating performance of the application layer, and micro
benchmark for testing the lower layers. We have implemented the smart contracts for all workloads for Ethereum, Parity
and Hyperledger, whose details are summarized in Table~\ref{tab:workloads}. Ethereum and Parity use the same execution
model, therefore they share the same smart contract implementations. 

\vspace{0.5cm}
\subsubsection{Macro benchmark workloads} We port two popular database benchmark workloads into \name, and three other
real
workloads found in the Ethereum blockchain. 

{\bf Key-value storage.} We implement a simple smart contract which functions as a key-value storage.
The {\tt WorkloadClient} is based on the YCSB driver~\cite{ycsb}. It preloads each store with a number of records, and
supports requests with different ratios of read and write operations. YCSB is widely used for evaluating NoSQL
databases.   

{\bf OLTP (Smallbank).} Unlike YCSB which does not consider transactions, Smallbank~\cite{smallbank} is a popular
benchmark for OLTP workload. It consists of three tables and four basic procedures simulating basic operations on bank
accounts. We implement it as a smart contract which simply transfers money from one account to another.  

{\bf EtherId.} This is a popular contract that implements a domain name registrar. It supports creation, modification
and ownership transfer of domain names. A user can request an existing domain by paying a certain amount to the current
domain's owner. This contract has been written for Ethereum blockchain, and can be ported to Parity without change. In
Hyperledger, we create two different key-value namespaces in the contract: one for storing the domain name data
structures, and another for users' account balances. In domain creation, the contract simply inserts domain value into
the first name space, using the domain name as the key. For ownership transfer, it checks the second namespace if the
requester has sufficient fund before updating the first namespace. To simulate real workloads, the contract contains a
function to pre-allocate user accounts with certain balances.   

{\bf Doubler.} This is a contract that implements a pyramid scheme. As shown in Figure~\ref{fig:doubler},
participants send money to this contract, and get rewards as more people join the scheme. In addition to the list of
participants and their contributions, the contract needs to keep the index of the next payout and updates the balance
accordingly after paying early participants. Similar to EthereId, this contract has already been written for Ethereum,
and can be ported to Parity directly. To implement it in Hyperledger, we need to translate the list operations into
key-value semantics, making the chaincode more bulky than the Ethereum counterpart. 

{\bf WavesPresale.} This contract supports digital token sales. It maintains two states: the total
number of tokens sold so far, and the list of previous sale transactions. It supports operations to add a new sale, to
transfer ownership of a previous sale, and to query a specific sale records. Ethereum and Parity support composite
structure data types, making it straightforward to implement the application logic. In contrast, in Hyperledger, we have
to translate this structure into key-value semantics by using separate key-value namespaces.   

\subsubsection{Micro benchmark workloads } The previous workloads test the performance of blockchain as a whole. As
discussed early in this section, a blockchain system comprises multiple layers, and each layer may have different
impact on the overall performance. We design several workloads to stress the layers in order to understand their
individual performance.  

{\bf DoNothing.} This contract accepts transaction as input and simply returns. In other words, it involves minimal number
of operations at the execution layer and data model layer, thus the overall performance will be mainly determined by the
consensus layer. Previous works on performance of blockchain consensus protocol~\cite{byzcoin, hybridconsensus} use {\em
time to consensus} to measure its performance. In \name, this metric is directly reflected in the transaction latency.  

{\bf Analytics.} This workload considers the performance of blockchain system in answering analytical queries about the
historical data. Similar to an OLAP benchmark,  this workload evaluates how the system implements scan-like and aggregate queries,
which are determined by its data model. Specifically, we implement two queries for extracting statistics from the
blockchain data: 
\begin{itemize}
\item [Q1:] {\em Compute the total transaction values committed between block i and block j}. 
\item [Q2:] {\em Compute the largest transaction value involving a given state (account) between block i and block
j}.  
\end{itemize}
In {\tt ClientWorkload}, we pre-load the blockhain with transactions carrying integer values (representing money
transferring) and the states with integer values. For Ethereum, both queries can be implemented via JSON-RPC APIs
that return transaction details and account balances at a specific block. For Hyperledger, however, the
second query must be implemented via a chaincode (VersionKVStore), because the system does not have APIs to query
historical states.  

{\bf IOHeavy.} Current blockchain systems rely on key-value storage to persist blockchain transactions and states. Each
storage system may perform differently under different workloads~\cite{zhang15}. This workload is designed to evaluate the
IO performance by invoking a contract that performs a large number of random writes and random reads to the contract's
states. The I/O bandwidth can be estimated via the observed transaction latency. 

{\bf CPUHeavy.} This workload measures the efficiency of the execution layer for computationally heavy tasks. EVM may be
fast at executing Ethereum specific operations, but it is unclear how it performs on general tasks for which
machine native codes may be more efficient. We deploy a smart contract which initializes a large array, and runs the quick
sort algorithm over it. The execution layer performance can then be measured by the observed transaction latency. 

\section{Performance Benchmark}
\label{sec:evaluation}
We selected Ethereum, Parity and Hyperledger for our study, as they occupy different positions in the blockchain design
space, and also for their codebase maturity. We evaluate the three systems using both macro and micro benchmark
workloads described in the previous section\footnote{We have released \name\ for public use~\cite{blockbench}.}. Our main findings are:

\begin{itemize} 
\item Hyperledger performs consistently better than Ethereum and Parity across the benchmarks. But it fails to scale up
to more than $16$ nodes. 
\item Ethereum and Parity are more resilient to node failures, but they are vulnerable to security attacks that forks
the blockchain. 
\item The main bottlenecks in Hyperledger and Ethereum are the consensus protocols, but for Parity the bottleneck is
caused by transaction signing. 
\item Ethereum and Parity incur large overhead in terms of memory and disk usage. Their execution engine is also less
efficient than that of Hyperledger. 
\item Hyperledger's data model is low level, but its flexibility enables customized optimization for analytical queries
of the blockchain data.   \end{itemize}


We used the popular Go implementation of Ethereum, {\em geth v1.4.18}, the Parity release {\em v1.6.0} and the
Hyperledger Fabric release {\em v0.6.0-preview}. We set up a private testnet for Ethereum and Parity by defining a
genesis block and directly adding peers to the miner network. For Ethereum, we manually tuned the {\tt difficulty}
variable in the genesis block to ensure that miners do not diverge in large networks. For Parity, we set the {\tt
stepDuration} variable to 1. In both Ethereum and Parity, {\tt confirmationLength} is set to $5$ seconds. The default
batch size in Hyperledger is $500$. 

The experiments were run on a 48-node commodity cluster. Each node has an E5-1650 3.5GHz CPU, 32GB RAM, 2TB hard drive,
running Ubuntu 14.04 Trusty, and connected to the other nodes via 1GB switch. The results below are averaged over 5
independent runs. {For Ethereum, we reserved 8 cores out of the available 12 cores per machine, so that the periodical
polls from the client's driver process do not interfere with the mining process (which is CPU intensive).}


\subsection{Macro benchmarks}
This section discusses the performance of the blockchain systems at the application layer, by running them with the YCSB
and Smallbank benchmarks over multiple nodes. 

\subsubsection{Throughput and latency}
\begin{figure*}
\subfloat[Peak performance]{\includegraphics[width=0.5\textwidth]{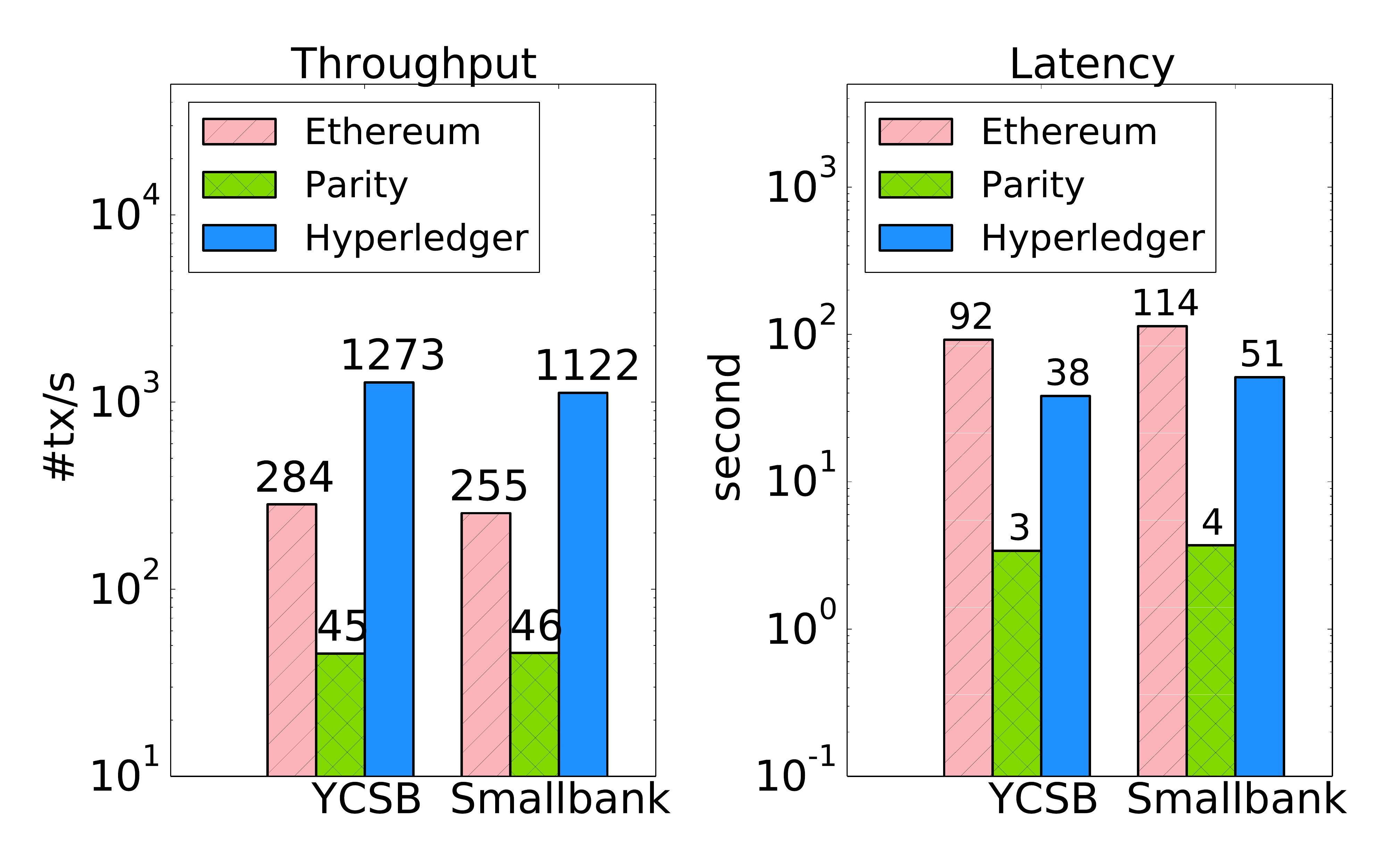}}
\subfloat[Performance with varying request rates]{\includegraphics[width=0.5\textwidth]{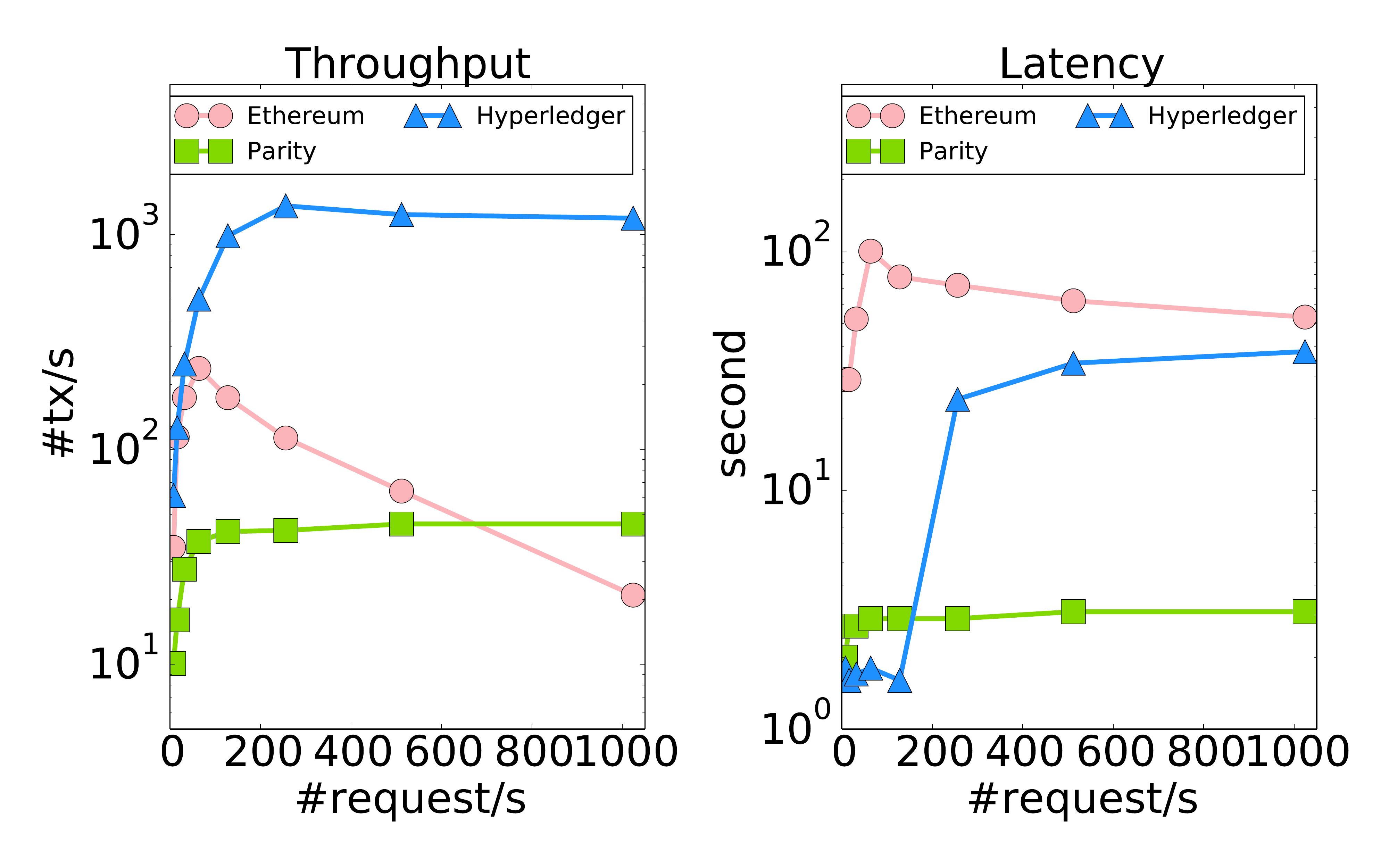}}
\caption{Blockchain performance with 8 clients and 8 servers.}
\label{fig:saturation}
\end{figure*}

We measured peak performance of the three systems with 8 servers and 8 concurrent clients over the period of 5 minutes.
Each client sends transactions to a server with a request rate varying from $8$ tx/s to $1024$ tx/s.
Figure~\ref{fig:saturation} shows the throughput and latency at peak, and how these metrics change with varying
transaction rates. 

We observe that in terms of throughput, Hyperledger outperforms other systems in both benchmarks.  Specifically, it has
up to $5.5$x and $28$x higher throughput than Ethereum and Parity respectively. Parity has the lowest latency and
Ethereum has the highest. The gap between Hyperledger and Ethereum is because of the difference in consensus protocol:
one is based on PBFT while the other is based on PoW. We measured CPU and network utilization during the experiments,
and observe that Hyperledger is communication bound whereas Ethereum is CPU bound (see Appendix B).  At 8 servers,
communication cost in broadcasting messages is much cheaper than block mining whose difficulty is set at roughly $2.5s$
per block. 


\begin{figure}
\centering
\includegraphics[width=0.5\textwidth]{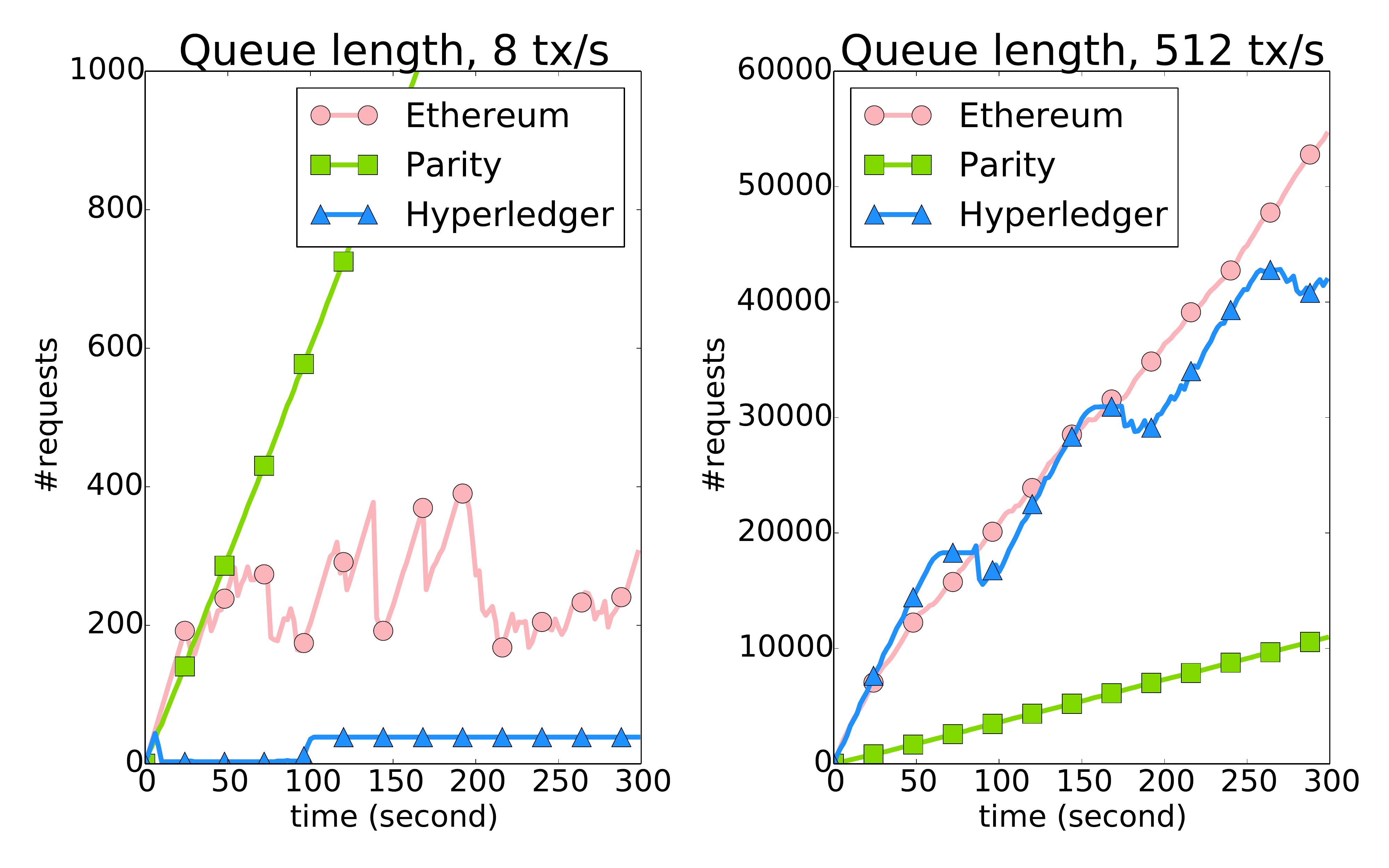}
\caption{Client's request queue, for request rates of 8 tx/s and 512 tx/s.}
\label{fig:queue_size}
\end{figure}

The performance gap between Parity and Hyperledger is not because of the consensus protocol, as we expect Parity's PoA protocol to
be simpler and more efficient than both PoW and PBFT (indeed, we observe that Parity has the same CPU utilization and
lower network utilization than Hyperledger). Figure~\ref{fig:saturation}[b,c] shows that Parity's throughput and latency
remains constant with increasing transaction rates (beyond $40$ tx/s). To understand its performance further, we
measure the queue of pending transactions at the client. { Figure~\ref{fig:queue_size} compares the queue sizes before and
after the systems reach their peak throughput. With only 8 tx/s, the queues for Ethereum and Hyperledger remain at
roughly constant sizes, but Parity's queue size increases as time passes. More interestingly, under high loads ($512$ tx/s
per client), Parity's queue is always smaller than Ethereum's and Hyperledger's.} This behavior indicates that Parity
processes transactions at a constant rate, and that it enforces a maximum client request rate at around $80$ tx/s. As a
consequence,
Parity achieves both lower throughput and latency than other systems. 

Another observation is that there are differences between YCSB and Smallbank workloads in Hyperledger and Ethereum.
There is a drop of $10\%$ in throughput and $20\%$ increase in latency. Since executing a Smallbank smart contract is
more expensive than executing a YCSB contract (there are more reading and writing to the blockchain's states), the
results suggest that there are non-negligible costs in the execution layer of blockchains.  

At its peak throughput, Hyperledger generates $3.1$ blocks per second and achieves the overall throughput of $1273$
tx/s. We remark that this throughput is far lower than what an in-memory database system can deliver (see Appendix B).
As the throughput is a function of the block sizes and block generation rate, we measured the effect of
increasing the block sizes in the three systems. The results (see Appendix B) demonstrate that with bigger block
sizes, the block generation rate decreases proportionally, thus the overall throughput does not improve.

\subsubsection{Scalability}
\begin{figure}
\centering
\includegraphics[width=0.5\textwidth]{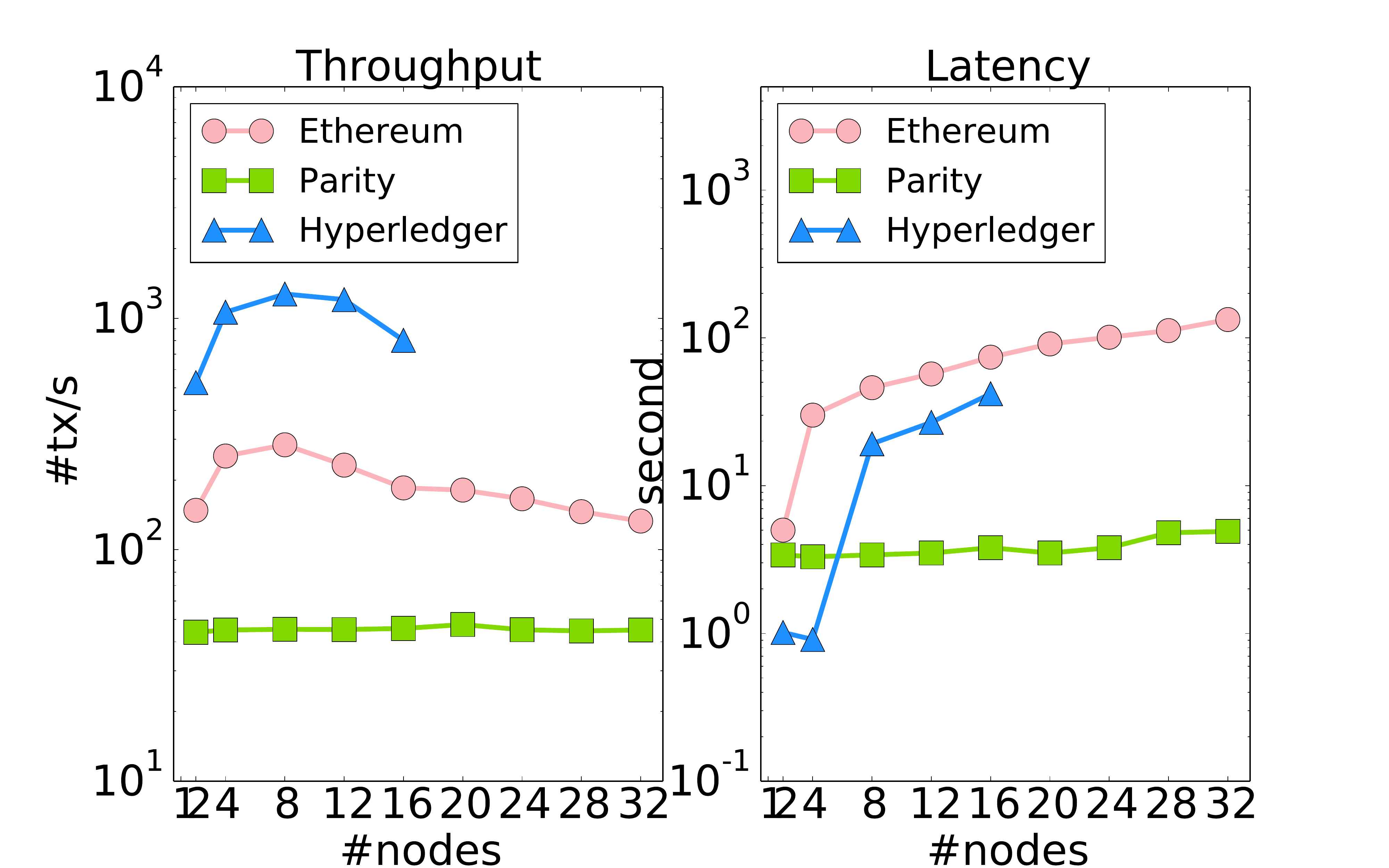}
\caption{Performance scalability (with the same number of clients and servers).}
\label{fig:scale}
\end{figure}

\begin{figure}
\centering
\includegraphics[width=0.45\textwidth]{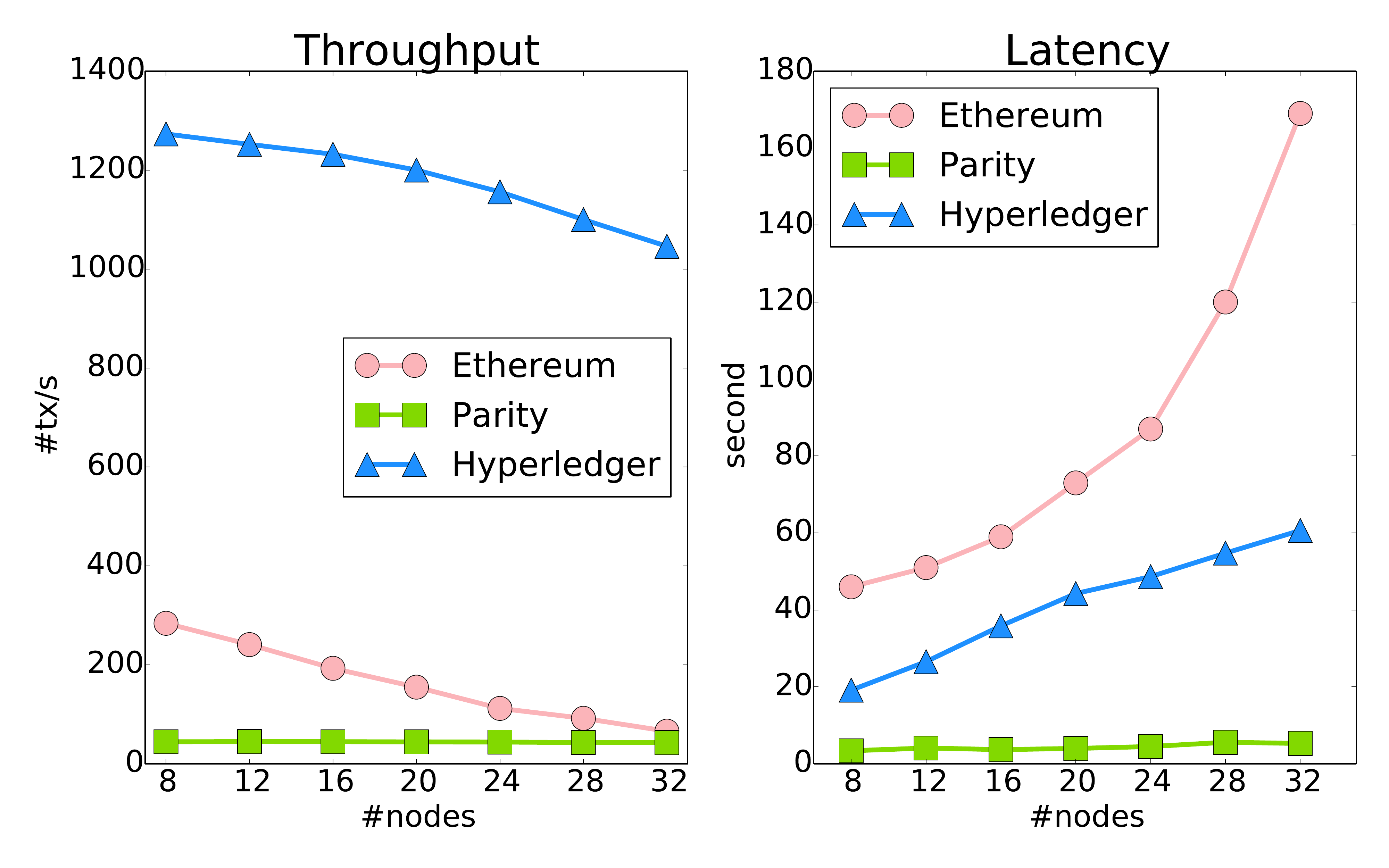}
\caption{Performance scalability (with 8 clients).}
\label{fig:scale_fixed}
\end{figure}

We fixed the client request rate and increased both the number of clients and the number of servers.
Figure~\ref{fig:scale} illustrates how well the three systems scale to handle larger YCSB workloads (the results for
Smallbank are similar and included in Appendix B).  Parity's performance remains constant as the network size
and offered load increase, due to the constant transaction processing rate at the servers.  Interestingly, while
Ethereum's throughput and latency degrade almost linearly beyond 8 servers, Hyperledger stops working beyond 16 servers.

To understand why Hyperledger failed to scale beyond 16 servers and 16 clients, we examined the system's logs and found
that the nodes were repeatedly trying and failing to reach consensus on new views which contain batches of transactions.
In fact, the servers were in different views and consequently were receiving conflicting view change messages from the
rest of the network. Further investigation reveals that conflicting views occurred because the consensus messages are
rejected by other peers on account of the message channel being full. As messages are dropped, the views start to
diverge and lead to unreachable consensus. In fact, we also observe that as time passes, client requests took longer to
return (see Appendix B), suggesting that the servers were over saturated in processing network messages.  We note,
however, that the original PBFT protocol guarantees both liveness and safety, thus Hyperledger's failure to scale beyond
16 servers is due to the implementation of the protocol.  In fact, in the latest codebase (which was updated after we
have finished our benchmark), the PBFT component was replaced by another implementation. We plan to evaluate this new
version in the future work.    

The results so far indicate that scaling both the number of clients and number of servers degrades the performance
and even causes Hyperledger to fail. We next examined the costs of increasing the number of servers alone while fixing
the number of clients. Figure~\ref{fig:scale_fixed} shows that the performance becomes worse as there are more servers,
meaning that the systems incur some network overheads. Because Hyperledger is communication bound, having more servers
means more messages being exchanged and higher overheads. For Ethereum, even though it is computation bound, it still
consumes a modest amount of network resources for propagating transactions and blocks to other nodes. { Furthermore, with
larger network, the difficulty is increased to account for the longer propagation delays. We observe that to prevent
the network from diverging, the difficulty level increases at higher rate than the number of nodes. Thus, one reason for
Ethereum's throughput degradation is due to network sizes.  Another reason is that in our settings, $8$ clients send
requests to only $8$ servers, but these servers do not always broadcast transactions to each other (they keep mining on
their own transaction pool). As a result, the network mining capability is not fully utilized. }

\subsubsection{Fault tolerance and security}
\begin{figure}
\centering
\includegraphics[width=0.4\textwidth]{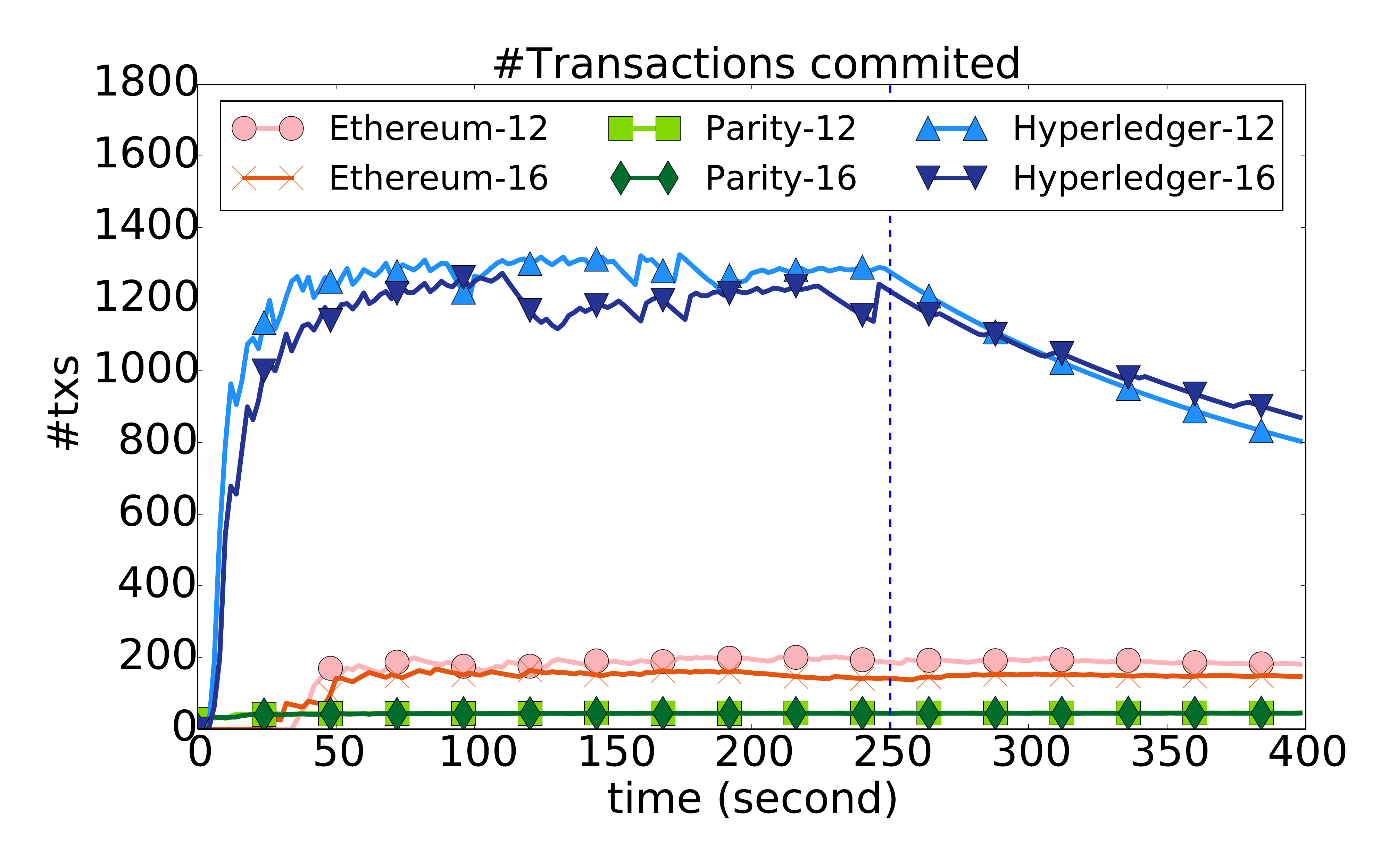}
\caption{Failing 4 nodes at $250^{th}$ second (fixed 8 clients) for 12 and 16 servers. {\em X-12} and {\em X-16} mean running 12 and 16 servers using blockchain {\em X} respectively.}
\label{fig:ft}
\end{figure}
\begin{figure}
\centering
\includegraphics[width=0.4\textwidth]{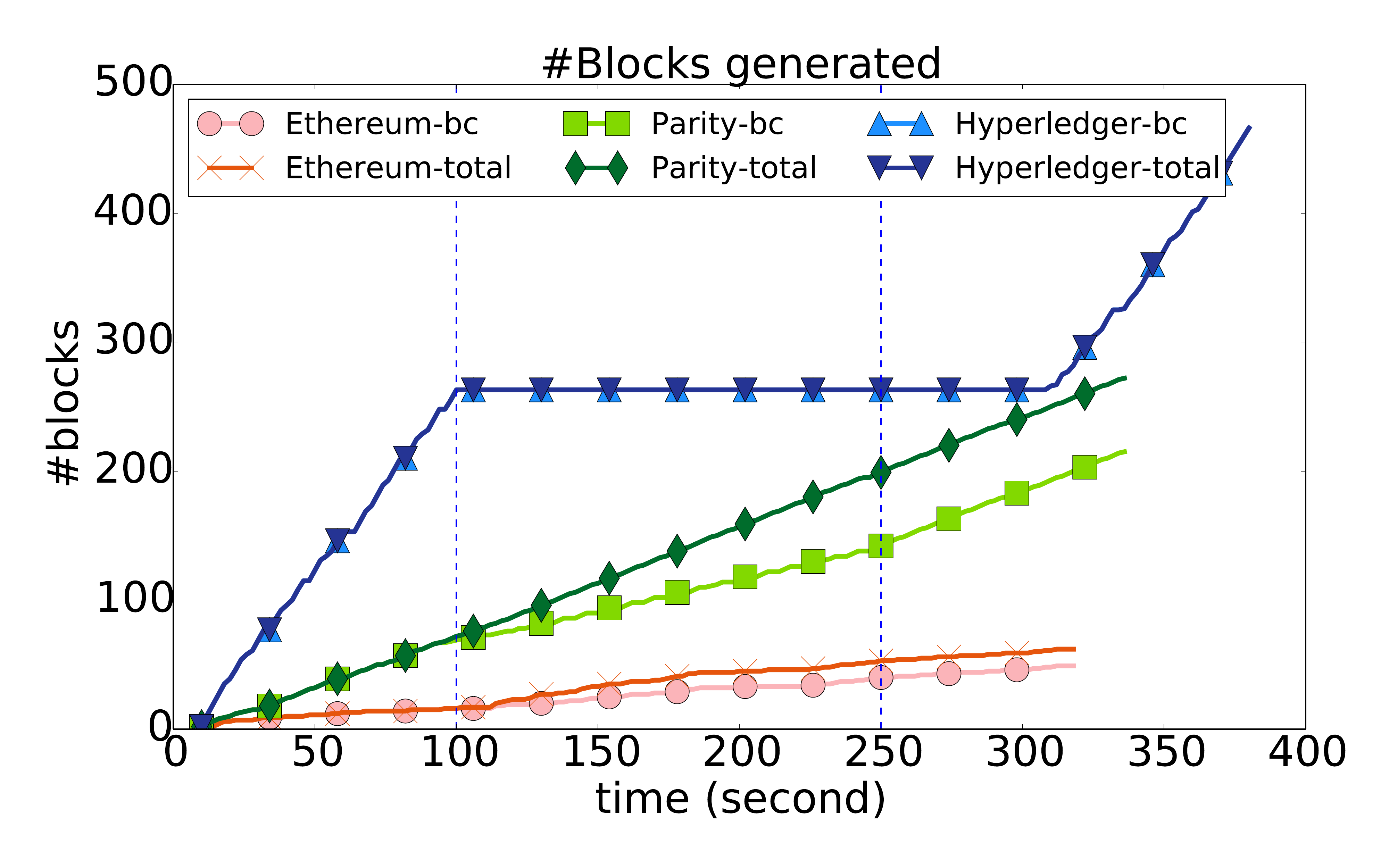}
\caption{Blockchain forks caused by attacks that partitions the network in half at $100^{\text{th}}$ second and lasts for
$150$ seconds. {\em X-total} means the total number of blocks generated in blockchain {\em X},
{\em X-bc} means the total number of blocks that reach consensus in blockchain {\em X}.} \label{fig:security}
\end{figure}

To evaluate how resilient the systems are to failures by crashing, we ran the systems with 8 clients for over 5 minutes,
during which we killed off 4 servers at $250^{th}$ second. Figure~\ref{fig:ft} shows that Ethereum is
nearly unaffected by the change, suggesting that the failed servers do not contributing significantly to the mining
process. In Parity, each node generates blocks at a constant rate, thus failing $4$ nodes means the remaining nodes are
given more time to generate more blocks, therefore the overall throughput is unaffected.  In contrast, the throughput
drops considerably in Hyperledger. For 12 servers, Hyperledger stops generating blocks after the failure, which is as
expected because the PBFT can only tolerate fewer than 4 failures in a 12-server network. With 16 servers, the system
still generated blocks but at a lower rate, which were caused by the remaining servers having to stabilize the network
after the failures by synchronizing their views.  

{We next simulated the attack that renders the blockchain vulnerable to double spending. The attack, described in
Section~\ref{subsec:metrics}, partitioned the network at $100^{th}$ second and lasted for $150$ seconds. We set the
partition size to be half of the original\footnote{We note that partitioning a $N$-node network in half does
not mean there are $N/2$ Byzantine nodes. In fact, Byzantine tolerance protocols do not count network adversary as
Byzantine failure}. Figure~\ref{fig:security} compares the vulnerability of the three systems running with $8$ clients
and $8$ servers. Recall that vulnerability is measured as the differences in the number of total blocks and the number of
blocks on the main branch (Section~\ref{subsec:metrics}), we refer to this as $\Delta$. Both Ethereum and Parity
blockchains fork at $100^{th}$ seconds, and $\Delta$ increases as time passes. For the attack duration, upto $30\%$ of the
blocks are generated in the forked branch, meaning that the systems are highly exposed to double spending or selfish
mining attacks. When the partition heals, the nodes come to consensus on the main branch and discard the forked blocks.
As a consequence, $\Delta$ stops increasing shortly after $250^{th}$ second. Hyperledger, in stark contrast, has no fork
which is as expected because its consensus protocol is proven to guaranteed safety. We note, however, that Hyperledger
takes longer than the other two systems to recover from the attacks (about $50$ seconds more). This is because of the
synchronization protocol executed after the partitioned nodes reconnect. }

\subsection{Micro benchmarks}
This section discusses the performance of the blockchain system at execution, data and consensus layers by evaluating
them with micro benchmark workloads. For the first two layers, the workloads were run using one client and one server.
For the consensus layer, we used 8 clients and 8 servers.  


\begin{figure}
\centering
\includegraphics[width=0.45\textwidth]{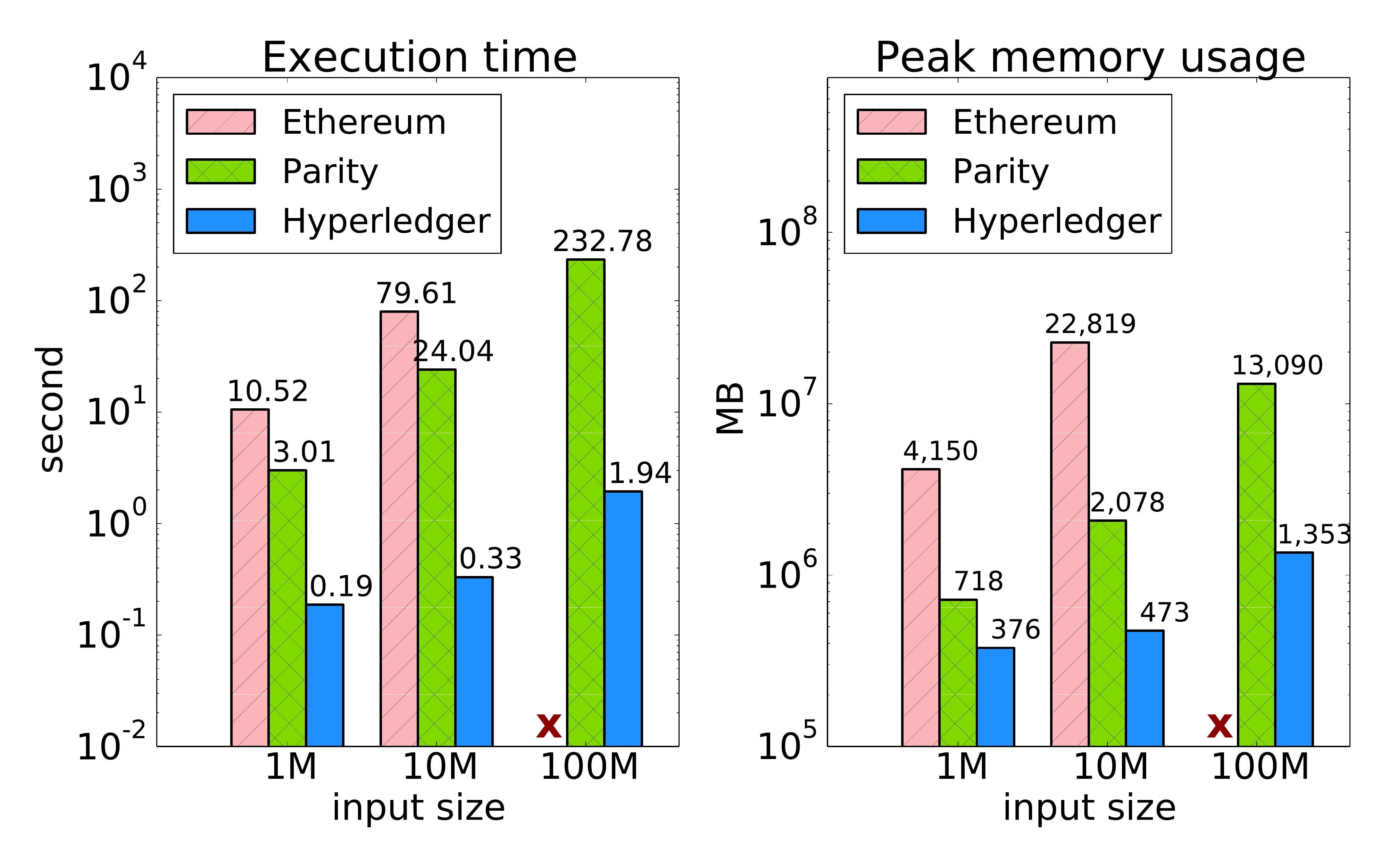}
\caption{CPUHeavy workload, {\bf `X'} indicates Out-of-Memory error.}
\label{fig:cpuheavy}
\end{figure}

\subsubsection{Execution layer}
{
We deployed the CPUHeavy smart contract that is initialized with an integer array of a given size. The array is
initialized in descending order. We invoked the contract to sort the array using quicksort algorithm, and measured the
execution time and server's peak memory usage. The results for varying input sizes are shown in
Figure~\ref{fig:cpuheavy}.  Although Ethereum and Parity use the same execution engine, i.e. EVM, Parity's
implementation is more optimized, therefore it is more computation and memory efficient.  An interesting finding is that
Ethereum incurs large memory overhead. In sorting $10M$ elements, it uses $22$GB of memory, as compared to $473$MB used
by Hyperledger. Ethereum runs out of memory when sorting more than $10$M elements.  In Hyperledger, the smart contract
is compiled and runs directly on the native machine within Docker environment, thus it does not have the overheads
associated with executing high-level EVM byte code. As the result, Hyperledger is much more efficient in term of speed
and memory usage.  Finally, we note that all three systems fail to make use of the multi-core architecture, i.e. they
execute the contracts using only one core.}

\begin{figure*}
\centering
\subfloat[Write]{\includegraphics[width=0.31\textwidth]{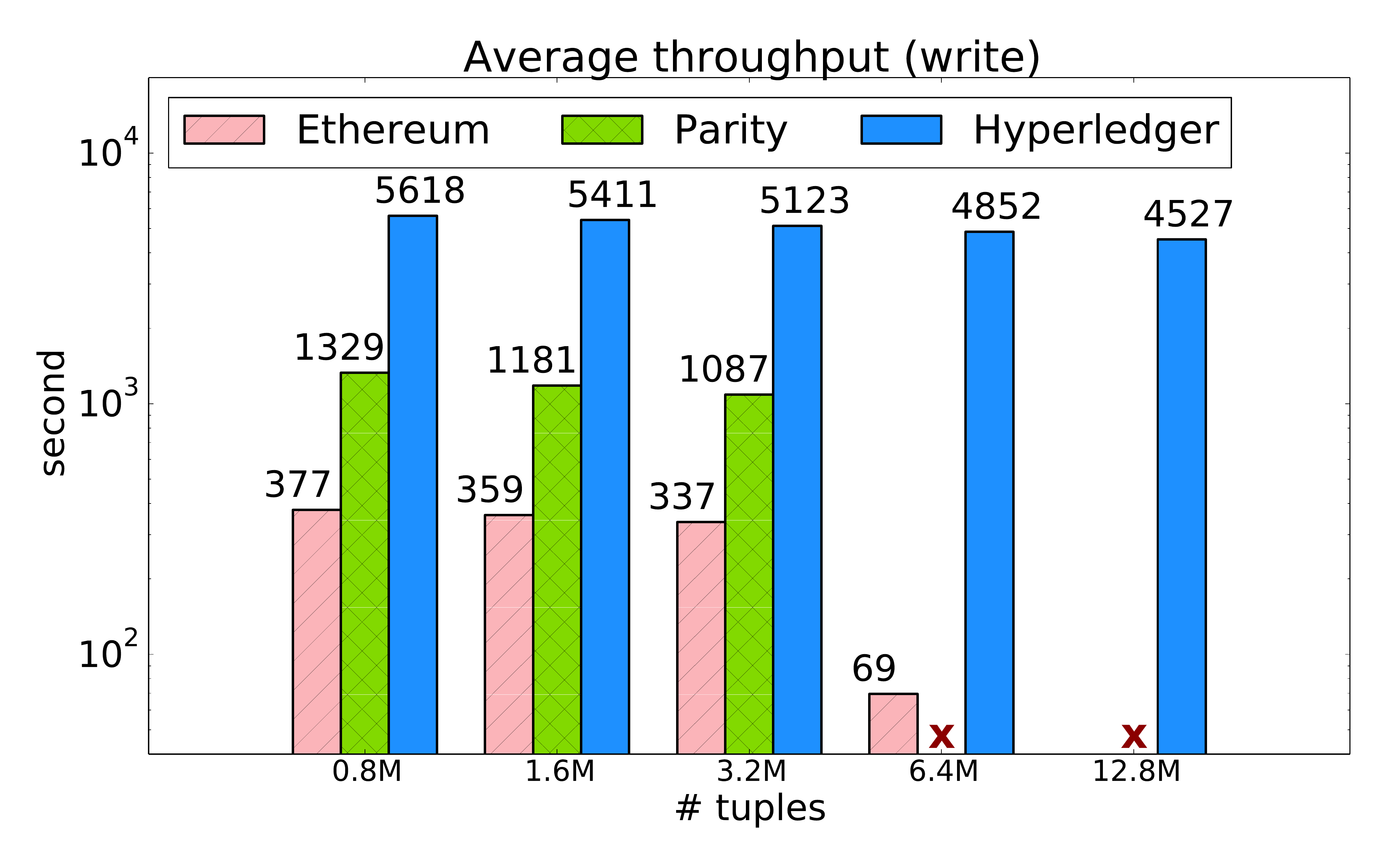}}
\subfloat[Read]{\includegraphics[width=0.31\textwidth]{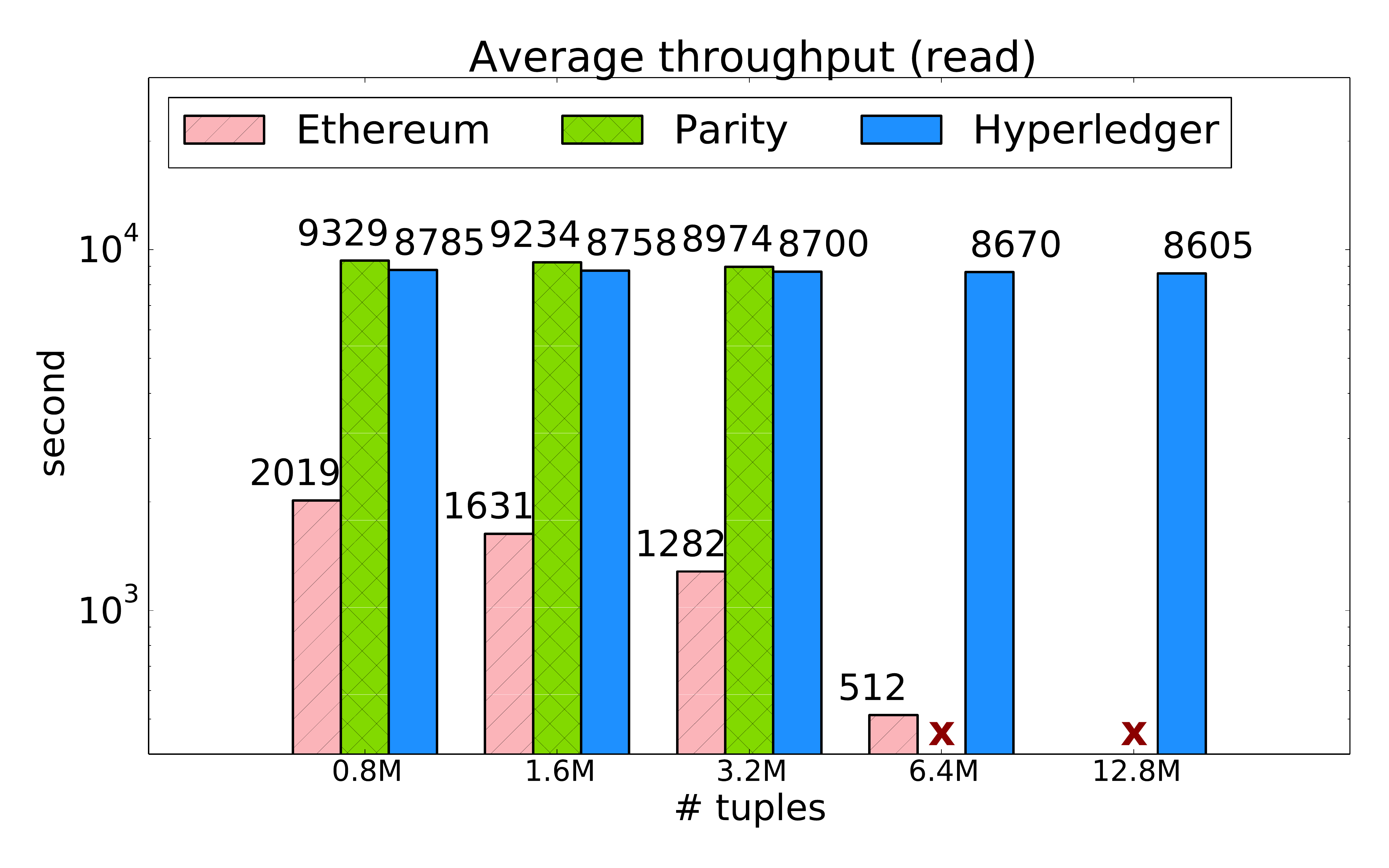}}
\subfloat[Disk usage]{\includegraphics[width=0.31\textwidth]{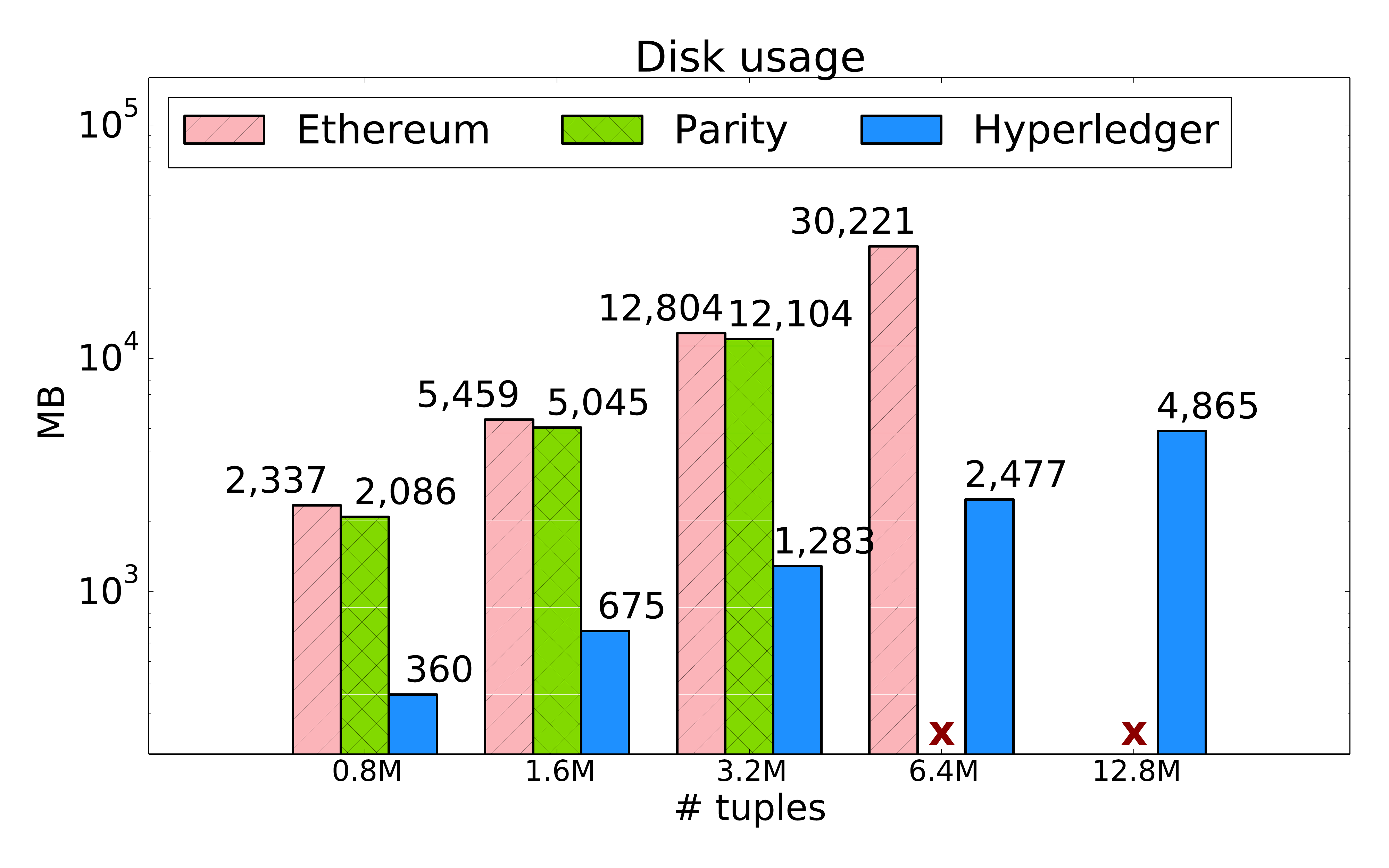}}
\caption{IOHeavy workload, {\bf `X'} indicates Out-of-Memory error.}
\label{fig:ioheavy}
\end{figure*}
\subsubsection{Data model}
\begin{figure*}
\centering
\subfloat[Analytics workload (Q1)]{\includegraphics[width=0.31\textwidth]{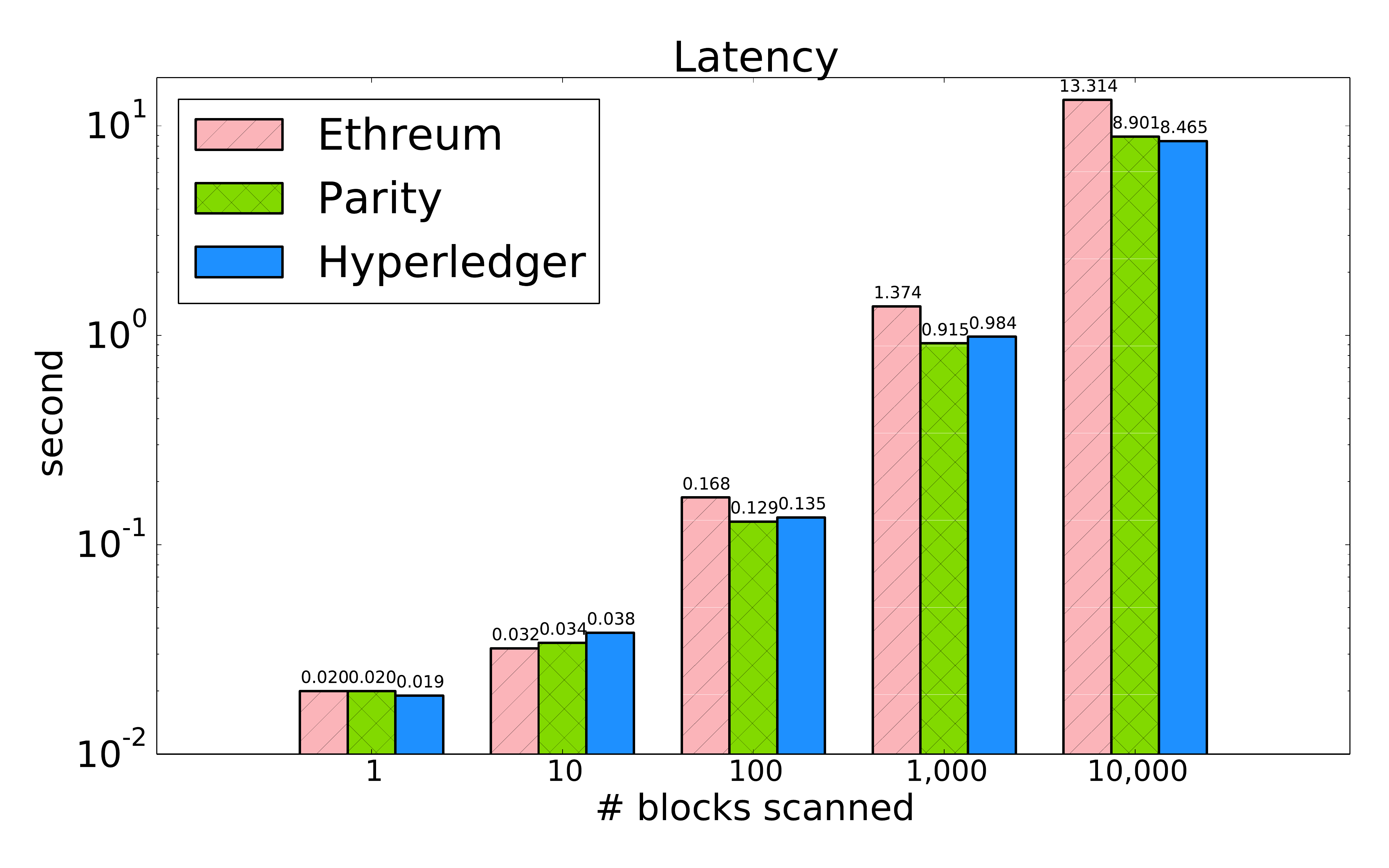}}
\subfloat[Analytics workload (Q2)]{\includegraphics[width=0.31\textwidth]{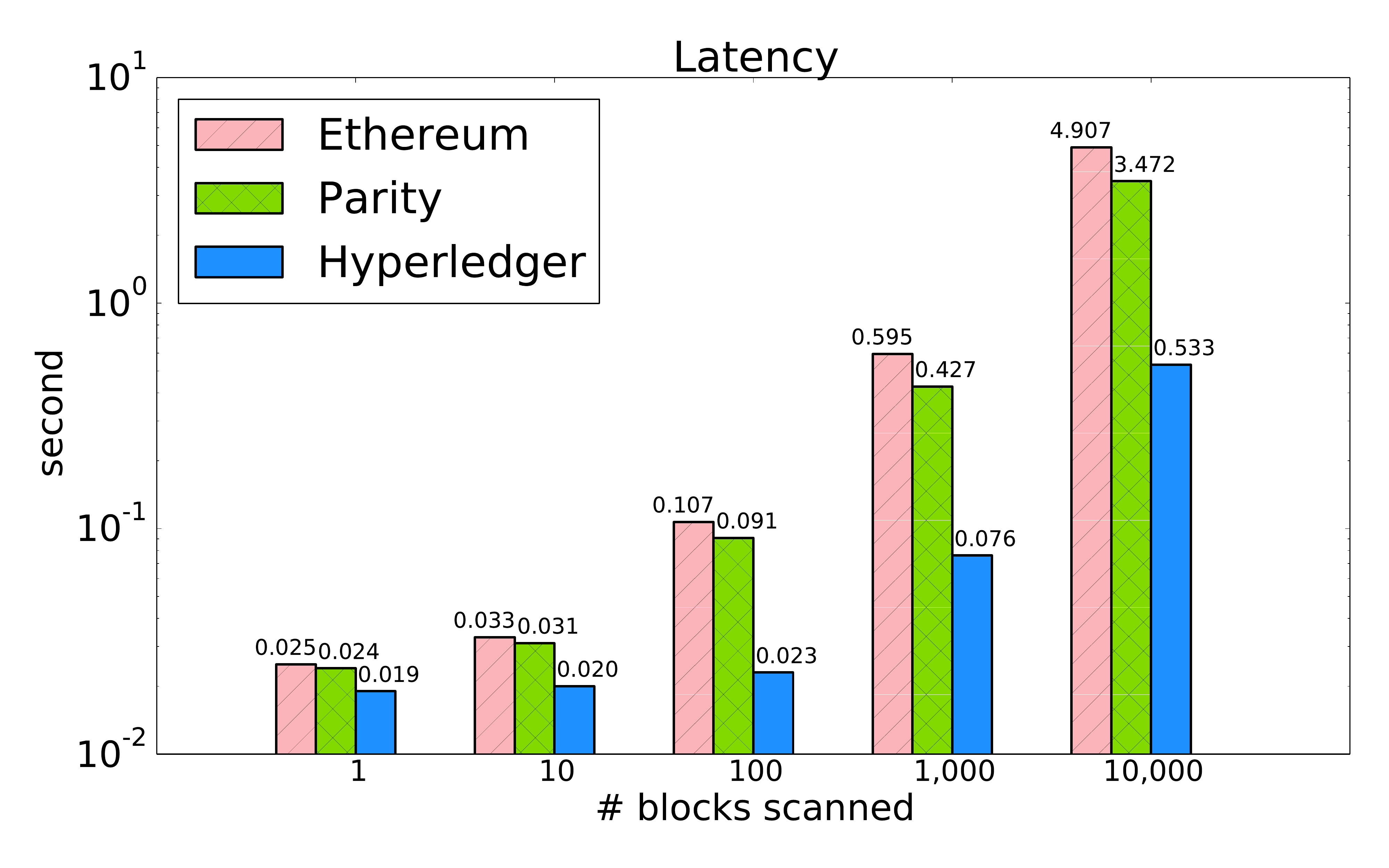}}
\subfloat[DoNothing workload]{\includegraphics[width=0.31\textwidth]{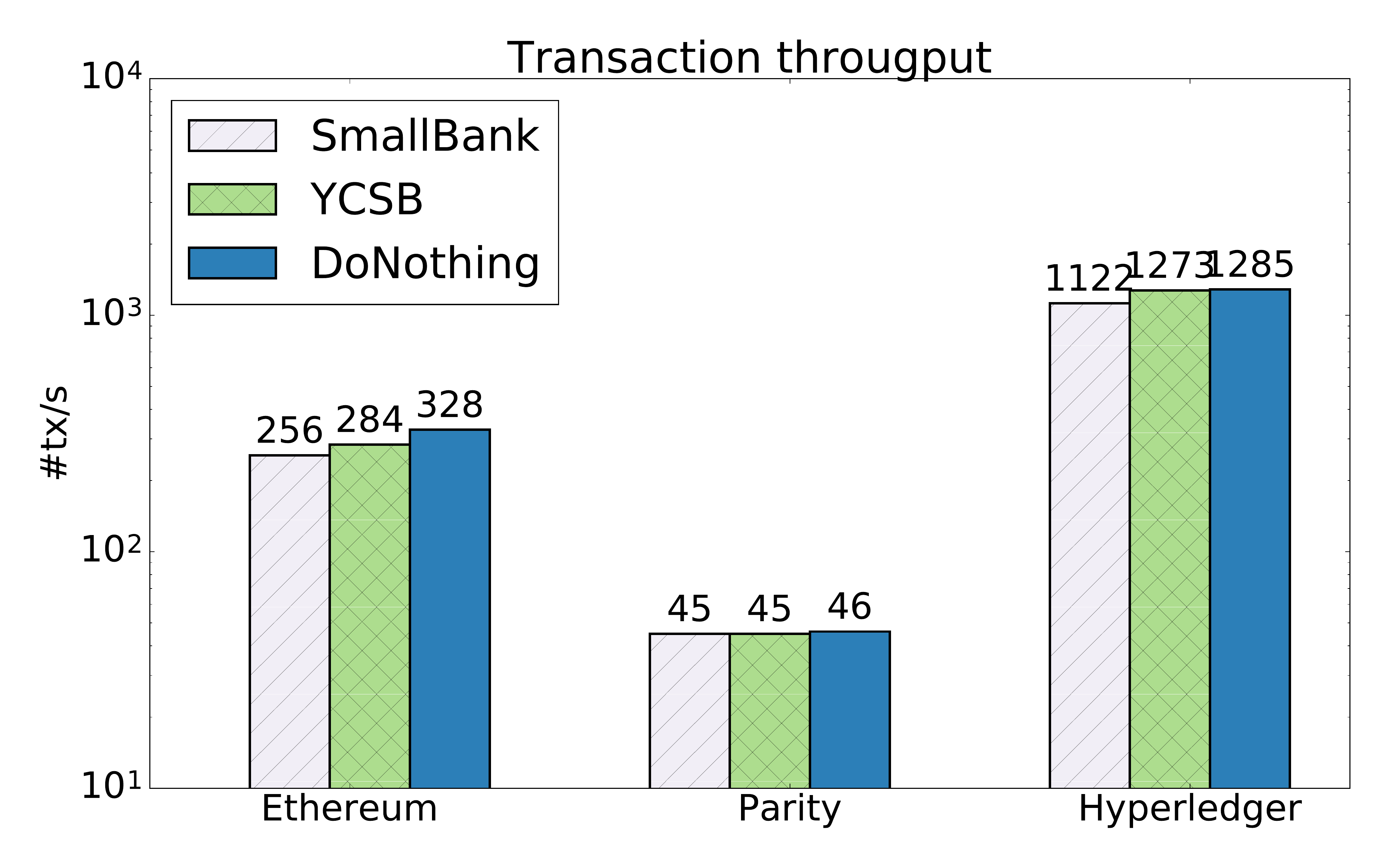}}
\caption{Analytics and DoNothing workloads.}
\label{fig:analytics}
\end{figure*}
{\textbf{IO Heavy.} We deployed the IOHeavy smart contract that performs a number of read and write operations of
key-value tuples. We used 20-byte keys and 100-byte values. Figure~\ref{fig:ioheavy} reports the throughput and disk
usage for these operations. Ethereum and Parity use the same data model and internal index structure, therefore they
incur similar space overheads. Both use an order of magnitude more storage space than Hyperledger which employs a simple
key-value data model. Parity holds all the state information in memory, so it has better I/O performance but fails to
handle large data (capped by over 3M states under our hardware settings). On the contrary, Ethereum only caches only
parts of the state in memory (using LRU for eviction policy), therefore it can handle more data than Parity at
the cost of throughput. Hyperledger leverages RocksDB to manage its states, which makes it more efficient at scale. }

\textbf{Analytic Queries.} We implemented the analytics workload by initializing the three systems with over $120,000$ accounts
with a fixed balance. We then pre-loaded them with $100,000$ blocks, each contains 3 transactions on average. The
transaction transfers a value from one random account to another random account. Due to Parity's overheads in signing
transactions when there are many accounts, we considered transactions using only $1024$ accounts. We then executed the
two queries described in Section~\ref{subsec:workloads} and measured their latencies.  Figure~\ref{fig:analytics} shows
that the performance for Q1 is similar, whereas Q2 sees a significant gap between Hyperledger and the rest. We note that
the main bottleneck for both Q1 and Q2 is the number of network (RPC) requests sent by the client. For Q1, the client
sends the same number of requests to all systems, therefore their performance are similar. On the other hand, for Q2 the
client sends one RPC per block to Ethereum and Parity, but only one RPC to Hyperledger because of our customized smart
contract implementation (see Appendix C). This saving in network roundtrip time translates to over $10$x improvement
in Q2 latency.  

\subsubsection{Consensus}
We deployed the DoNothing smart contract that accepts a transaction and returns immediately. We measured the throughput
of this workload and compare against that of YCSB and Smallbank. The differences compared to other workloads, shown in
Figure~\ref{fig:analytics}[c] is indicative of the cost of consensus protocol versus the rest of the software stack. In
particular, for Ethereum we observe $10\%$ increases in throughput as compared to YCSB, which means that execution of
the YCSB transaction accounts for the $10\%$ overhead. We observe no differences among these workloads in Parity,
because the bottleneck in Parity is due to transaction signing (even empty transactions still need to be signed), not
due to consensus or transaction execution. 

\section{Discussion}
\label{sec:discussion}
{\noindent\textbf{Understanding blockchain systems.} Our framework is designed to provide better understanding of the
performance and design of different private blockchain systems. As more and more blockchain systems are being proposed,
each offering different sets of feature, \name's main value is that it narrows down the design space into four distinct
abstraction layers. Our survey of current blockchain systems (see Appendix A) show that the four layers are
sufficient to capture the key characteristics of these systems. By benchmarking these layers, one can gain insights into
the design trade-offs and performance bottlenecks. In this paper, for example, by running the IOHeavy workload we
identify that Parity trades performance for scalability by keeping states in memory. Another example is the trade-off
in data model made by Hyperledger. On the one hand, the simple key-value model means some analytical queries cannot be
directly supported. On the other hand, it enables optimization that helps answering the queries more efficiently.
Finally, we identify that the bottleneck in Parity is not due to the consensus protocol, but due to the server's transaction
signing. We argue that such insights are not easy to extract without a systematic analysis framework.   }


\noindent\textbf{Usability of blockchain.} Our experience in working with the three blockchain systems confirms the belief
that in its current state blockchain are not yet ready for mass usage. Both their designs and codebases are still being
refined constantly, and there are no other established applications beside crypto-currency. Of the three systems,
Ethereum is more mature both in terms of its codebase, user base and developer community. Another usability issue we
encountered is in porting smart contracts from one system to another, because of their distinct programming models (see
Section~\ref{sec:design}). This is likely to be exacerbated as more blockchain platforms are being
proposed~\cite{ripple,crypti}.     

\noindent\textbf{Bringing database designs into blockchain.} The challenge in scaling blockchain by improving its consensus
protocols is being addressed in many recent works~\cite{byzcoin,elastico}. However, as we demonstrated in the previous section,
there are other performance bottlenecks. We propose four approaches in applying design principles from database
systems to improve blockchain. 

{\em Decouple storage, execution engine and consensus layer from each other, then optimize and scale them
independently.} For instance, current systems employ generic key-value storage, which may not be best suited to the
unique data structure and operations in blockchain. UStore~\cite{ustore} demonstrates that a storage designed
around the blockchain data structure is able to achieve better performance than existing implementations.

{\em Embrace new hardware primitives.} Many data processing systems are taking advantage of new hardware to boost their
performance~\cite{in-memory,zhang15,farm}. For blockchain, using trusted hardware, the underlying Byzantine fault tolerance
protocols can be modified to incur fewer network messages~\cite{a2m}. Systems like Parity and Ethereum can take
advantage of multi-core CPUs and large memory to improve contract execution and I/O performance.  

{\em Sharding.} Blockchain is essentially a replicated state machine system, in which each
node maintains the same data. As such, blockchains are fundamentally different to database systems such as
H-Store in which the data is partitioned (or sharded) across the nodes. Sharding helps reduce the computation
cost and can make transaction processing faster. The main challenge with sharding is to ensure consistency
among multiple shards. However, existing consistency protocols used in database systems do not work under
Byzantine failure. Nevertheless, their designs can offer insights into realizing a more scalable sharding
protocol for blockchain. Recent work~\cite{elastico} has demonstrated the feasibility of sharding the
consensus protocol, making important steps towards partitioning the entire blockchain.

{\em Support declarative language.} Having a set of high-level operations that can be composed in a declarative manner
makes it easy to define complex smart contracts. It also opens up opportunities for low-level optimizations that speed
up contract execution. 

\section{Related Work}
\label{sec:related}
Performance studies of blockchain systems have so far been restricted to public blockchains. For example,
\cite{wattenhofer12,croman2016scaling} analyze the effect of block sizes and network propagation time on the overall
throughputs. Recent proposals for improving Bitcoin
performance~\cite{gervais16,byzcoin,elastico,bitcoin-ng,hybridconsensus} have mainly focused on the consensus layer,
in which analytical models or network simulations are used to validate the new designs. Various aspects of Ethereum,
such as their block processing time (for syncing with other nodes) and transactions processing time, have also been
benchmarked~\cite{ethbench,ethcore}. Our analysis using \name\ differs from these works in that it is the first to
evaluate private blockchains systems at scale against database workloads. Furthermore, it compares two different systems
and analyzes how their designs affect the overall performances. Future extensions of \name\ would enable more
comparative evaluations of the key components in blockchain.  

There are many standard frameworks for benchmarking database systems. OLTP-Bench~\cite{oltpbench} contains standard
workloads such as TPC-C for transactional systems. YCSB~\cite{ycsb} contains key-value workloads.
HiBench~\cite{hibench} and BigBench~\cite{bigbench} feature big-data analytics workloads for MapReduce-like systems.
\name\ shares the same high-level design as these frameworks, but its workloads and main driver are designed
specifically for blockchain systems.   

\section{Conclusion}
\label{sec:conclusion}
In this paper we proposed the first benchmarking framework, called \name, for evaluating private blockchain systems.
\name\ contains workloads for measuring the data processing performance, and workloads for
understanding the performance at different layers of the blockchain. Using \name, we conducted comprehensive analysis of
three major blockchain systems, namely Ethereum, Parity and Hyperledger with two macro benchmarks and four micro
benchmarks. The results showed that current blockchains are not well suited for large scale data processing workloads.
 We demonstrated several bottlenecks and design trade-offs at
different layers of the software stack.  
\newpage
\section*{Acknowledgment}
We would like to thank the anonymous reviewers for their comments and suggestions that help us improve the
paper. Special thanks to Hao Zhang, Loi Luu, the developers from Ethereum, Parity and Hyperledger projects for
helping us with the experiment setup. This work is funded by the National Research Foundation, Prime
Minister's Office, Singapore, under its Competitive Research Programme (CRP Award No. NRF-CRP8-2011-08).
\balance
\bibliographystyle{abbrv}
\bibliography{ref}

\begin{thebibliography}{10}

\bibitem{blockbench}
{BlockBench:} private blockchains benchmarking.
\newblock \url{https://github.com/ooibc88/blockbench}.

\bibitem{ethereum}
Ethereum blockchain app platform.
\newblock \url{https://www.ethereum.org/}.

\bibitem{watsoniot}
Ibm watson iot.
\newblock \url{http://www.ibm.com/internet-of-things}.

\bibitem{leveldb}
Leveldb.
\newblock \url{https://leveldb.org}.

\bibitem{monax}
Monax: The ecosystem application platform.
\newblock \url{https://monax.io}.

\bibitem{rocksdb}
Rocksdb.
\newblock \url{https://rocksdb.org}.

\bibitem{bgp_hijack}
M.~Apostolaki, A.~Zohar, and L.~Vanbever.
\newblock Hijacking bitcoin: Large-scale network attacks on crypto-currencies.
\newblock \url{https://arxiv.org/abs/1605.07524}, 2016.

\bibitem{bailis14}
P.~Bailis, A.~Fekete, M.~J. Franklin, A.~Ghodsi, J.~M. Hellerstein, and
  I.~Stoica.
\newblock Coordination avoidance in database systems.
\newblock In {\em VLDB}.

\bibitem{bonneau2015sok}
J.~Bonneau, A.~Miller, J.~Clark, A.~Narayanan, J.~A. Kroll, and E.~W. Felten.
\newblock Sok: Research perspectives and challenges for bitcoin and
  crypto-currencies.
\newblock In {\em 2015 IEEE Symposium on Security and Privacy}, pages 104--121.
  IEEE, 2015.

\bibitem{smallbank}
M.~Cahill, U.~Rohm, and A.~D. Fekete.
\newblock Serializable isolation for snapshot databases.
\newblock In {\em SIGMOD}, 2008.

\bibitem{castro1999practical}
M.~Castro and B.~Liskov.
\newblock Practical byzantine fault tolerance.
\newblock In {\em Proceedings of the third symposium on Operating systems
  design and implementation}, pages 173--186. USENIX Association, 1999.

\bibitem{a2m}
B.-G. Chun, P.~Maniatis, S.~Shenker, and J.~Kubiatowicz.
\newblock Attested append-only memory: Making adversaries stick to their word.
\newblock In {\em SOSP}, 2007.

\bibitem{ycsb}
B.~F. Cooper, A.~Silberstein, E.~Tam, R.~Ramakrishnan, and R.~Sears.
\newblock Benchmarking cloud serving systems with ycsb.
\newblock In {\em SoCC}, 2010.

\bibitem{spanner}
J.~C. Corbett and J.~D. et~al.
\newblock Spanner: Google's globally-distributed database.
\newblock In {\em OSDI}, 2012.

\bibitem{croman2016scaling}
K.~Croman, C.~Decker, I.~Eyal, A.~E. Gencer, A.~Juels, A.~Kosba, A.~Miller,
  P.~Saxena, E.~Shi, and E.~G{\"u}n.
\newblock On scaling decentralized blockchains.
\newblock In {\em Proc. 3rd Workshop on Bitcoin and Blockchain Research}, 2016.

\bibitem{crypti}
Crypti.
\newblock A decentralized application platform.
\newblock \url{https://crypti.me}.

\bibitem{wattenhofer12}
C.~Decker and R.~Wattenhofer.
\newblock Information propagation in bitcoin network.
\newblock In {\em P2P}, 2013.

\bibitem{oltpbench}
D.~E. Difallah, A.~Pavlo, C.~Curino, and P.~Cudre-Mauroux.
\newblock Oltp-bench: An extensible testbed for benchmarking relational
  databases.
\newblock In {\em VLDB}, 2013.

\bibitem{ustore}
A.~Dinh, J.~Wang, S.~Wang, W.-N. Chin, Q.~Lin, B.~C. Ooi, P.~Ruan, K.-L. Tan,
  Z.~Xie, H.~Zhang, and M.~Zhang.
\newblock {UStore:} a distributed storage with rich semantics.
\newblock \url{https://arxiv.org/pdf/1702.02799.pdf}.

\bibitem{sybil}
J.~Douceur.
\newblock The sybil attack.
\newblock In {\em IPTPS}, 2002.

\bibitem{farm}
A.~Dragojevic, D.~Narayanan, E.~B. Nightingale, M.~Renzelmann, A.~Shamis,
  A.~Badam, and M.~Castro.
\newblock No compromises: distributed transactions with consistency,
  availability and performance.
\newblock In {\em SOSP}, 2015.

\bibitem{parity}
Ethcore.
\newblock Parity: next generation ethereum browser.
\newblock \url{https://ethcore.io/parity.html}.

\bibitem{ethcore}
Ethcore.
\newblock Performance analysis.
\newblock \url{https://blog.ethcore.io/performance-analysis/}.

\bibitem{ethbench}
Ethereum.
\newblock Ethereum benchmarks.
\newblock \url{https://github.com/ethereum/wiki/wiki/Benchmarks}.

\bibitem{bitcoin-ng}
I.~Eyal, A.~E. Gencer, E.~G. Sirer, and R.~van Renesse.
\newblock Bitcoin-ng: A scalable blockchain protocol.
\newblock In {\em NSDI}, 2016.

\bibitem{eyal14}
I.~Eyal and E.~G. Sirer.
\newblock Majority is not enough: Bitcoin mining is vulnerable.
\newblock In {\em Fiancial Cryptography}, 2014.

\bibitem{gervais16}
A.~Gervais, G.~O. Karame, K.~Wust, V.~Glykantizis, H.~Ritzdorf, and S.~Capkun.
\newblock On the security and performance of proof of work blockchains.
\newblock \url{https://eprint.iacr.org/2016/555.pdf}.

\bibitem{bigbench}
A.~Ghazal, T.~Rabl, M.~Hu, F.~Raab, M.~Poess, A.~Crolotte, and H.-A. Jacobsen.
\newblock Bigbench: towards an industry standard benchmark for big data
  analytics.
\newblock In {\em SIGMOD}, 2013.

\bibitem{gs16}
G.~S. Group.
\newblock Blockchain: putting theory into practice, 2016.

\bibitem{eclipse}
E.~Heilman, A.~Kendler, A.~Zohar, and S.~Goldberg.
\newblock Eclipse attacks on {Bitcoin}'s peer-to-peer network.
\newblock In {\em USENIX Security}, 2015.

\bibitem{hyperledger}
Hyperledger.
\newblock Blockchain technologies for business.
\newblock \url{https://www.hyperledger.org}.

\bibitem{hibench}
Intel.
\newblock Hibench suite.
\newblock \url{https://github.com/intel-hadoop/HiBench}.

\bibitem{zab}
F.~P. Junqueira, B.~C. Reed, and M.~Serafini.
\newblock Zab: high-performance broadcast for primary-backup systems.
\newblock In {\em Dependable Systems and Networks}, 2011.

\bibitem{byzcoin}
E.~Kokoris-Kogias, P.~Jovanovic, N.~Gailly, I.~Khoffi, L.~Gasser, and B.~Ford.
\newblock Enhancing bitcoin security and performance with strong consistency
  via collective signing.
\newblock In {\em USENIX Security}, 2016.

\bibitem{paxos}
L.~Lamport.
\newblock Paxos made simple.
\newblock {\em SIGACT News}, 2001.

\bibitem{qian16}
Q.~Lin, P.~Chang, G.~Chen, B.~C. Ooi, K.-L. Tan, and Z.~Wang.
\newblock Towards a non-2pc transaction management in distrubted database
  systems.
\newblock In {\em SIGMOD}, 2016.

\bibitem{elastico}
L.~Luu, V.~Narayanan, C.~Zhang, K.~Baweija, S.~Gilbert, and P.~Saxena.
\newblock A secure sharding protocol for open blockchains.
\newblock In {\em CCS}, 2016.

\bibitem{Luu2015demystifying}
L.~Luu, J.~Teutsch, R.~Kulkarni, and P.~Saxena.
\newblock {Demystifying Incentives in the Consensus Computer}.
\newblock {\em CCS '15}, pages 706--719, 2015.

\bibitem{melonport}
Melonport.
\newblock Blockchain software for asset management.
\newblock \url{http://melonport.com}.

\bibitem{morgan16}
J.~Morgan and O.~Wyman.
\newblock Unlocking economic advantage with blockchain. a guide for asset
  managers., 2016.

\bibitem{nakamoto2008bitcoin}
S.~Nakamoto.
\newblock Bitcoin: A peer-to-peer electronic cash system, 2008.

\bibitem{raft}
D.~Ongaro and J.~Ousterhout.
\newblock In search of an understandable consensus algorithm.
\newblock In {\em USENIX ATC}, 2014.

\bibitem{hybridconsensus}
R.~Pass and E.~Shi.
\newblock Hybrid consensus: efficient consensus in the permissionless model.
\newblock \url{https://eprint.iacr.org/2016/917.pdf}.

\bibitem{ripple}
Ripple.
\newblock Ripple.
\newblock \url{https://ripple.com}.

\bibitem{ghost}
Y.~Sompolinsky and A.~Zohar.
\newblock Accelerating bitcoin's transaction processing: fast money grows on
  trees, not chains.
\newblock Cryptology ePrint Archive, Report 2013/881, 2013.
\newblock \url{https://eprint.iacr.org/2013/881.pdf}.

\bibitem{hstore}
M.~Stonebraker, S.~Madden, D.~J. Abadi, S.~Harizopoulos, N.~Hachem, and
  P.~Helland.
\newblock The end of and architectural era (it's time for a complete rewrite).
\newblock In {\em VLDB}, 2007.

\bibitem{in-memory}
K.-L. Tan, Q.~Cai, B.~C. Ooi, W.-F. Wong, C.~Yao, and H.~Zhang.
\newblock In-memory databases: Challenges and opportunities from software and
  hardware perspectives.
\newblock {\em SIGMOD Records}, 44(2), 2015.

\bibitem{thomson12}
A.~Thomson, T.~Diamond, S.~chun Weng, K.~Ren, P.~Shao, and D.~J. Abadi.
\newblock Calvin: fast distributed transaction for partitioned database
  systems.
\newblock In {\em SIGMOD}, 2012.

\bibitem{vulkolic15}
M.~Vukolic.
\newblock The quest for scalable blockchain fabric: proof-of-work vs. bft
  replication.
\newblock In {\em Open Problems in Network Security - iNetSec}, 2015.

\bibitem{zhang15}
H.~Zhang, G.~Chen, B.~C. Ooi, K.-L. Tan, and M.~Zhang.
\newblock In-memory big data management and processing: a survey.
\newblock {\em TKDE}, 2015.

\end{thebibliography}

\appendix
\label{sec:appx}
\vspace{0.5cm}

%

\section{Survey of Blockchain Platforms}

\begin{table*}
\caption{Comparison of blockchain platforms}
\label{table:statb}
\centering  
\begin{tabular}{m{1.8cm}<{\centering} | m{2.3cm}<{\centering} | m{2.3cm}<{\centering} | m{3.5cm}<{\centering} | m{2.1cm}<{\centering} | m{2.7cm}<{\centering}}
\Xhline{1.2pt}
\hline
\hline
& \bf{Application} & \bf{Smart contract execution} & \bf{Smart contract language} & \bf{Data model} & \bf{Consensus}\\
\hline
Hyperledger & Smart contract & Dockers & Golang, Java & Account-based & PBFT\\
\hline
Ethereum & Smart contract, Crypto-currency & EVM & Solidity, Serpent, LLL & Account-based & Ethash (PoW)\\
\hline
Eris-DB & Smart contract & EVM & Solidity & Account-based & Tendermint (BFT)\\
\hline
Ripple & Crypto-currency & - & - & UTXO-based & Ripple Consensus Ledger (PoS)\\
\hline
ScalableBFT & Smart contract & Haskell Execution & Pact & Account-based & ScalableBFT\\
\hline
Stellar & Smart contract & Dockers & JavaScript, Golang, Java, Ruby, Python, C\# & Account-based & Stellar Consensus Protocol\\
\hline
Dfinity & Smart contract & EVM & Solidity, Serpent, LLL & Account-based & Blockchain Nervous System\\
\hline
Parity & Smart contract & EVM & Solidity, Serpent, LLL & Account-based & Proof of Authority\\
\hline
Tezos & Smart contract, Crypto-currency & Dockers & Tezos Contract Script Language & Account-based & Proof of Stake\\
\hline
Corda & Smart contract & JVM & Kotlin, Java & UTXO-based & Raft\\
\hline
Sawtooth Lake & Smart contract & TEE & Python & Account-based & Proof of Elapsed Time\\\hline
\hline
\Xhline{1.2pt}
\end{tabular}
\end{table*}

  
We compare eleven promising blockchain platforms in Table \ref{table:statb}. We can see that all but Ripple support
smart contracts. Ethereum, Eris-DB, Dfinity and Parity execute the contracts using Ehtereum Virtual Machine (EVM),
whereas Corda runs them in Java Virtual Machine (JVM). Hyperledger, Stellar and Tezos employ Docker images, ScalableBFT
takes Haskell execution environment, and Sawtooth Lake launches contracts on top of Trusted Execution Environment (TEE)
such as Intel Software Guard Extensions (SGX). These platforms also support different languages to develop smart
contracts. For example, Solidity, Serpent and LLL are mainly used in Ethereum, Dfinity and Parity, while Eris-DB only
supports Solidity. Hyperledger, Stellar, Corda and Sawtooth Lake exploit various mature programming languages, such as
Python, Java, Golang, etc. ScalableBFT and Tezos even develop their own smart contract languages. Most blockchain
platforms' data models are account-based. Two exceptions in the table are Ripple and Corda. Their data models are similar
to Bitcoin's unspent transaction outputs (UTXO) which represents the coins in the network. 

Each platform offers different consensus protocols. Hyperledger implements PBFT in the version we evaluated, while
Ethereum implements a variation of PoW (Proof-of-Work). Eris-DB builds on top of Tendermint protocol but only works in
the latest version (v 0.12). Ripple and Tezos deploy Proof-of-Stake (PoS) schemes (the one in Ripple is referred to
Ripple Consensus Ledger) where the next block is created based on accounts' wealth, i.e., the stake. Parity takes
another consensus protocol, Proof-of-Authority (PoA), which holds a predefined set of "authorities" to create new
blocks in a fixed time slot and secure the blockchain network. Sawtooth Lake uses Proof-of-Elapsed-Time (PoET) as its consensus protocol, which in
nature is a lottery algorithm and decides the creator of block arbitrarily. Stellar develops its own mechanism, Stellar
Consensus Protocol, which is a construction for decentralized Byzantine agreement. There is no source code that helps
determine which consensus protocol Dfinity uses, but its documents suggest that a Blockchain Nervous System will govern
the whole platform via a voting mechanism based on {\em neurons} that interact with each other and are controlled by users.

\vspace{0.5cm}
\section{Macro Benchmarks}
\begin{figure}
\centering
\includegraphics[width=0.42\textwidth]{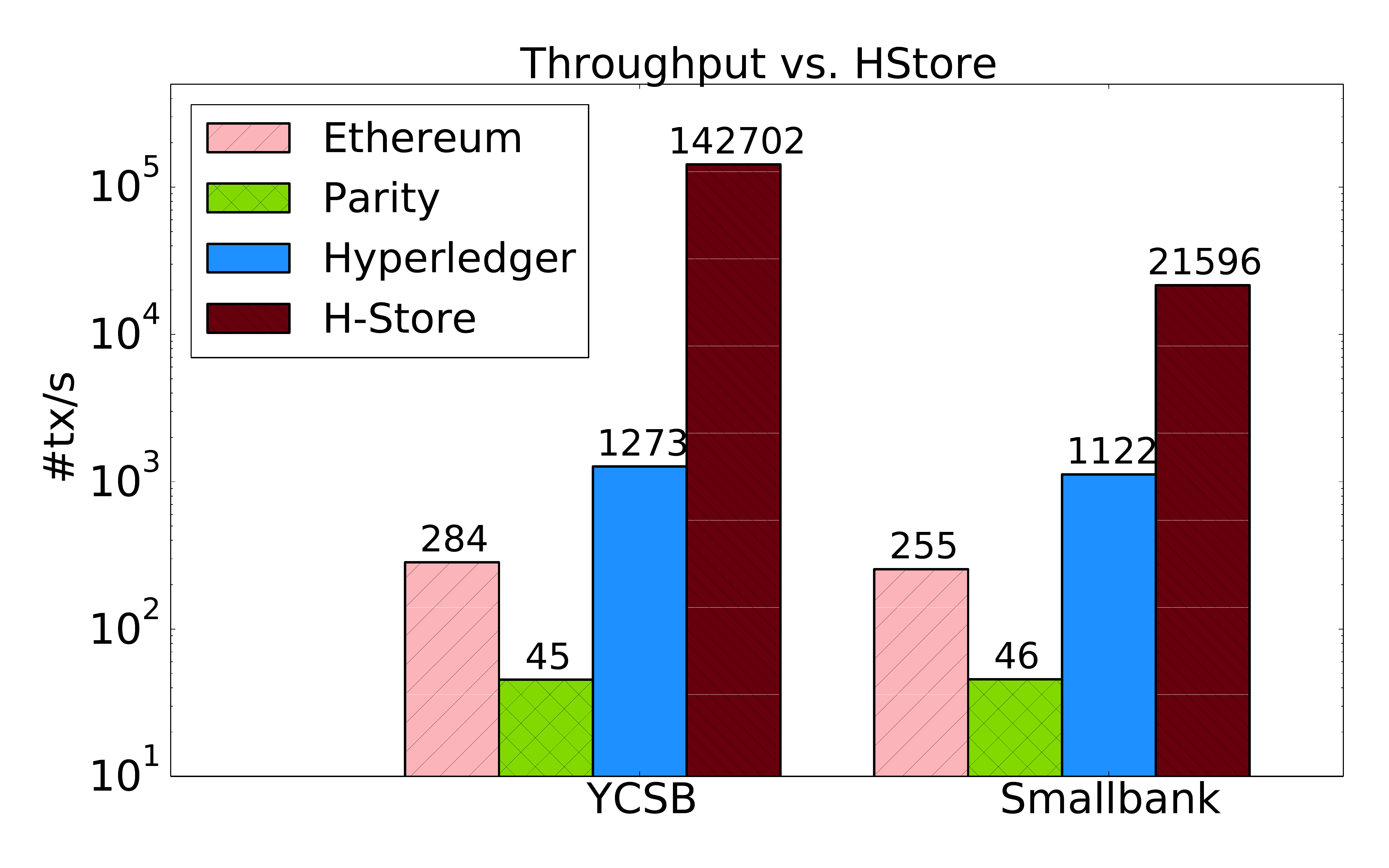}
\caption{Performance of the three blockchain systems versus H-Store.}
\label{fig:vs_hstore}
\end{figure}

We compared the performance of the three blockchain systems against a popular in-memory database system, namely H-Store,
using the YCSB and Smallbank workload. We ran H-Store's own benchmark driver and set the transaction rate at 100,000
tx/s. Figure~\ref{fig:vs_hstore} shows at least an order of magnitude gap in throughput and two order of magnitude in
latency. Specifically, H-Store achieves over 140K tx/s throughput while maintaining sub-millisecond latency.
The gap in performance is due to the cost of consensus protocols. For YCSB, for example, H-Store requires almost no
coordination among peers, whereas Ethereum and Hyperledger suffer the overhead of PoW and PBFT.  

An interesting observation is the overhead of Smallbank. Recall that Smallbank is a more complex transactional workload
than YCSB, in which multiple keys are updated in a single transaction. Smallbank is simple but is representative of the
large class of transactional workloads such as TPC-C. We observe that in H-Store, Smallbank achieves $6.6$x lower throughput
and $4$x higher latency than YCSB, which indicates the cost of distributed transaction management protocol, because
H-Store is a sharded database. In contrast, the blockchain suffers modest degradation in performance: $10\%$ in
throughput and $20\%$ in latency. This is because each node in blockchain maintains the entire state (replicated state
machine), thus there is no overhead in coordinating distributed transactions as the data is not partitioned.    

The results demonstrate that blockchain performs poorly at data processing tasks currently handled by database systems.
However, we stress that blockchains and databases are designed with different goals and assumptions. Specifically, the
protocols for Byzantine failure tolerance are an overkill for traditional database settings where there are only crash
failures. Other features which are optional in most database systems are cryptographic signatures on every single
transaction, and wide-area fully replicated state machines. Although databases are designed without security features
and tolerance to Byzantine failures, we remark that the gap remains too high for blockchains to be disruptive to
incumbent database systems. Nevertheless, the popularity of blockchain is a clear indication that there is a need for a
Byzantine tolerant data processing systems which can accommodate a large number of users. 

\begin{figure}
\centering
\includegraphics[width=0.42\textwidth]{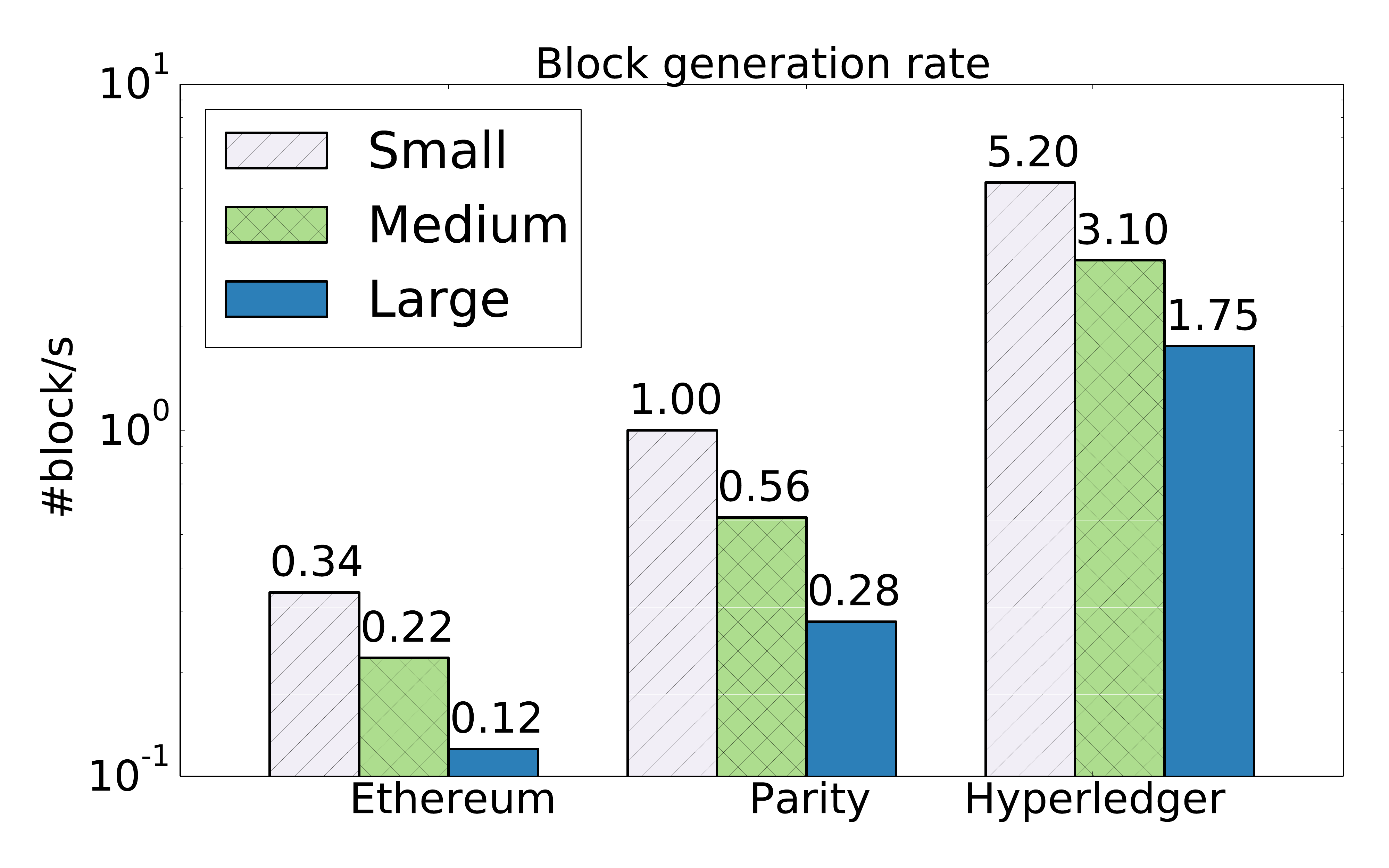}
\caption{Block generation rate.}
\label{fig:blockrate}
\end{figure}

Figure~\ref{fig:blockrate} shows the effect of varying block sizes in the overall throughput. While it is
straightforward to set the block size in Hyperledger by configuring the {\tt batchSize} variable, there is no direct way
to specify the same in Ethereum. An Ethereum miner uses {\tt gasLimit} value to restrict the overall cost in
constructing a block, thus we tuned this value to simulate different sizes. In Parity, {\tt gasLimit} is not applicable
to local transaction and it has no effect on the block size. Instead, we observe that the block size can be controlled
by tuning {\tt stepDuration} value, which essentially decides how much time a validator can use to build a block. In the
experiments, {\em medium} size refers to the default settings, whereas {\em large} and {\em small} refer to $2$x and
$0.5$x of the default size. The results show that increases in block sizes lead to proportional decreases in block
generation rate, meaning that the overall throughput does not improve.  

\begin{figure}
\centering
\includegraphics[width=0.42\textwidth]{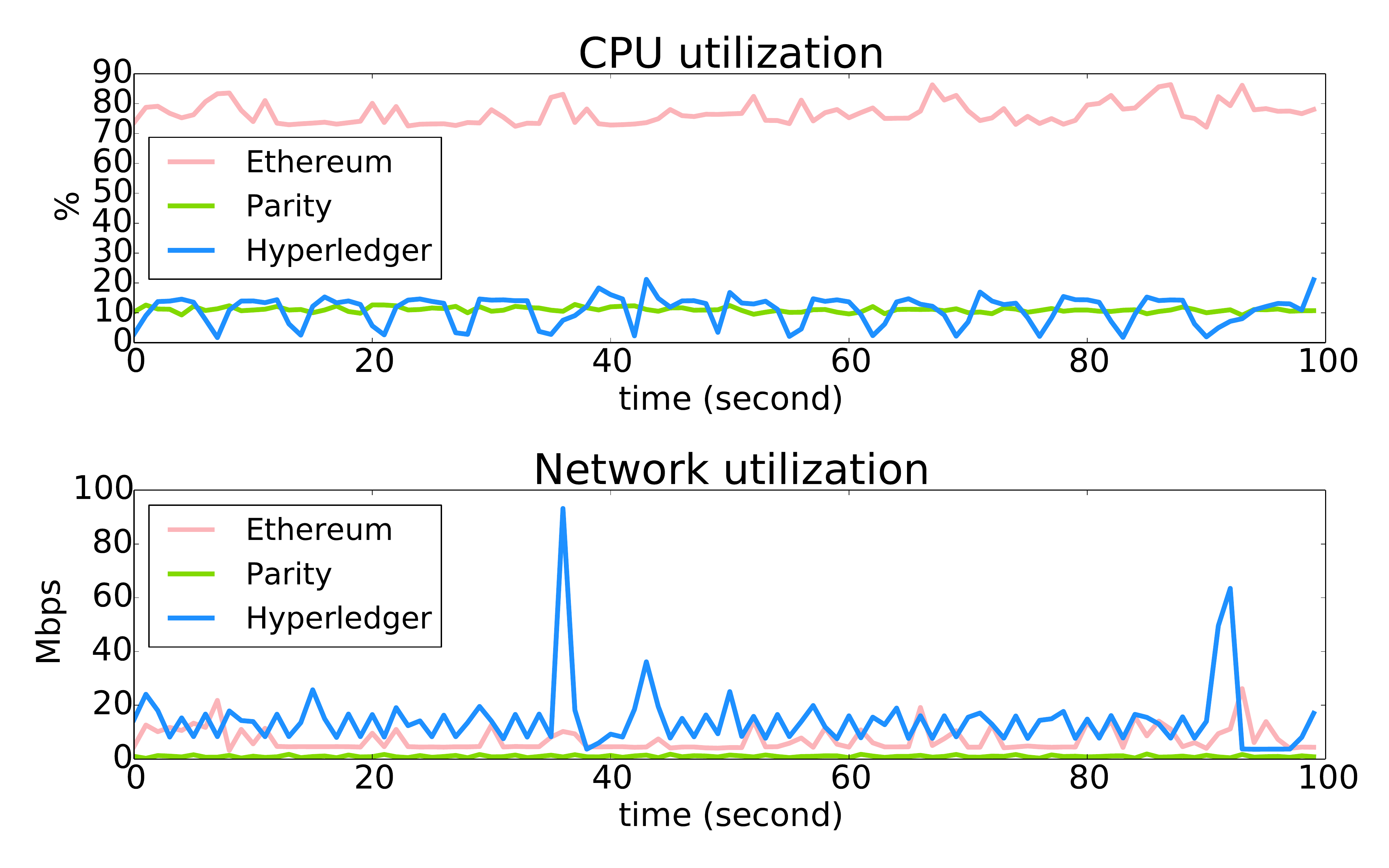}
\caption{Resource utilization.}
\label{fig:resources}
\end{figure}

Figure~\ref{fig:resources} compares CPU and network utilization of the three systems over the period of 100 seconds. It is
easy to see that Ethereum is CPU bound, as it fully utilizes 8 CPU cores. Hyperledger, on the other hand, uses CPU
sparingly and spends the rest of the time on network communication. Parity, in contrast, has lower resource footprints
than other two systems. For Ethereum and Hyperledger, the pattern is the direct consequence of the consensus protocol:
PoW is CPU bound whereas PBFT is communication bound. 

\begin{figure}
\centering
\includegraphics[width=0.4\textwidth]{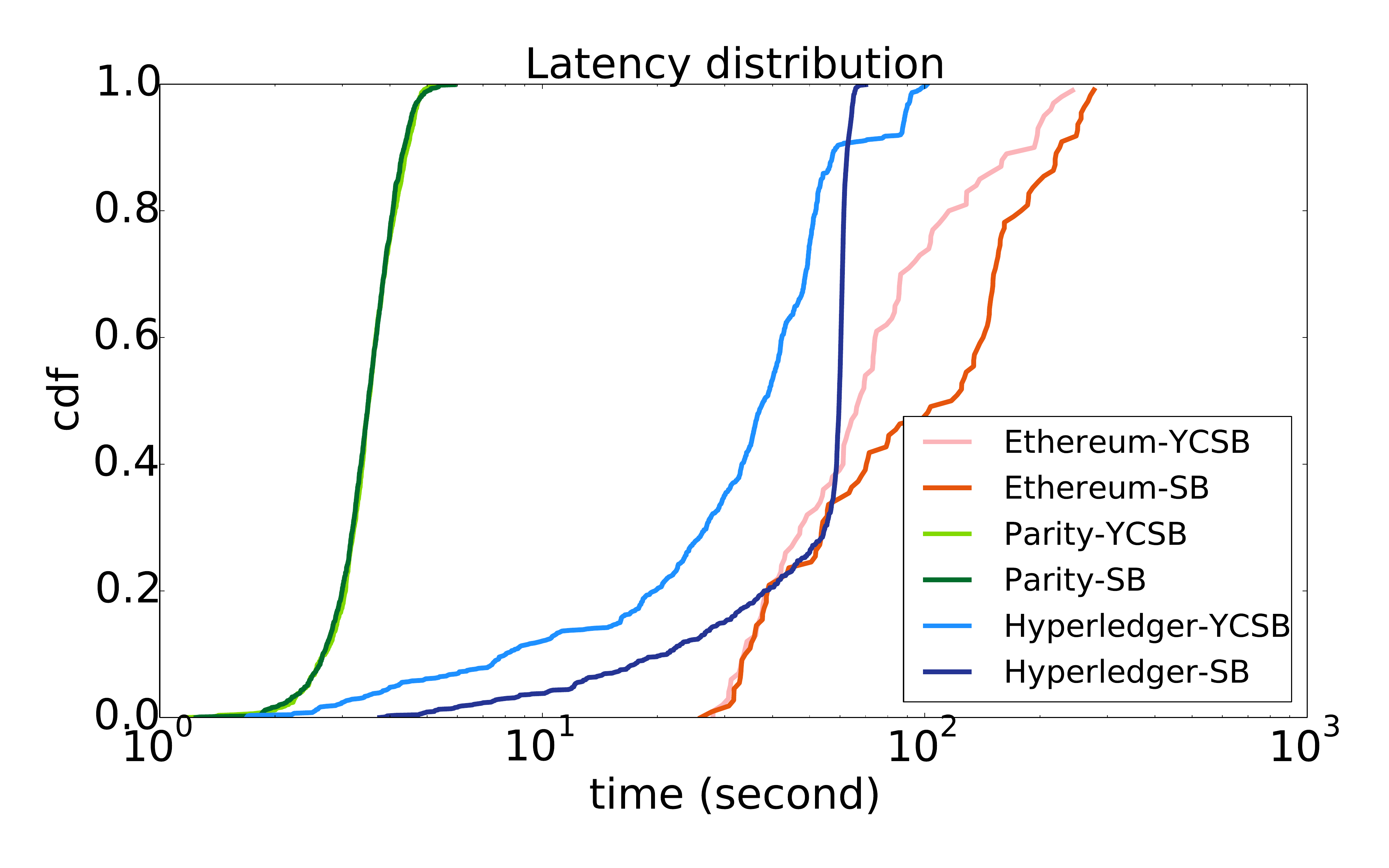}
\caption{Latency distribution.}
\label{fig:cdfs}
\end{figure}

Figure~\ref{fig:cdfs} shows the latency distribution. Ethereum has both higher latency and higher variance, because PoW
is a randomized process which means the duration between blocks are unpredictable. Parity has the lowest variance
because the server restricts the client request rate at $80$ tx/s.  

\begin{figure}
\centering
\includegraphics[width=0.42\textwidth]{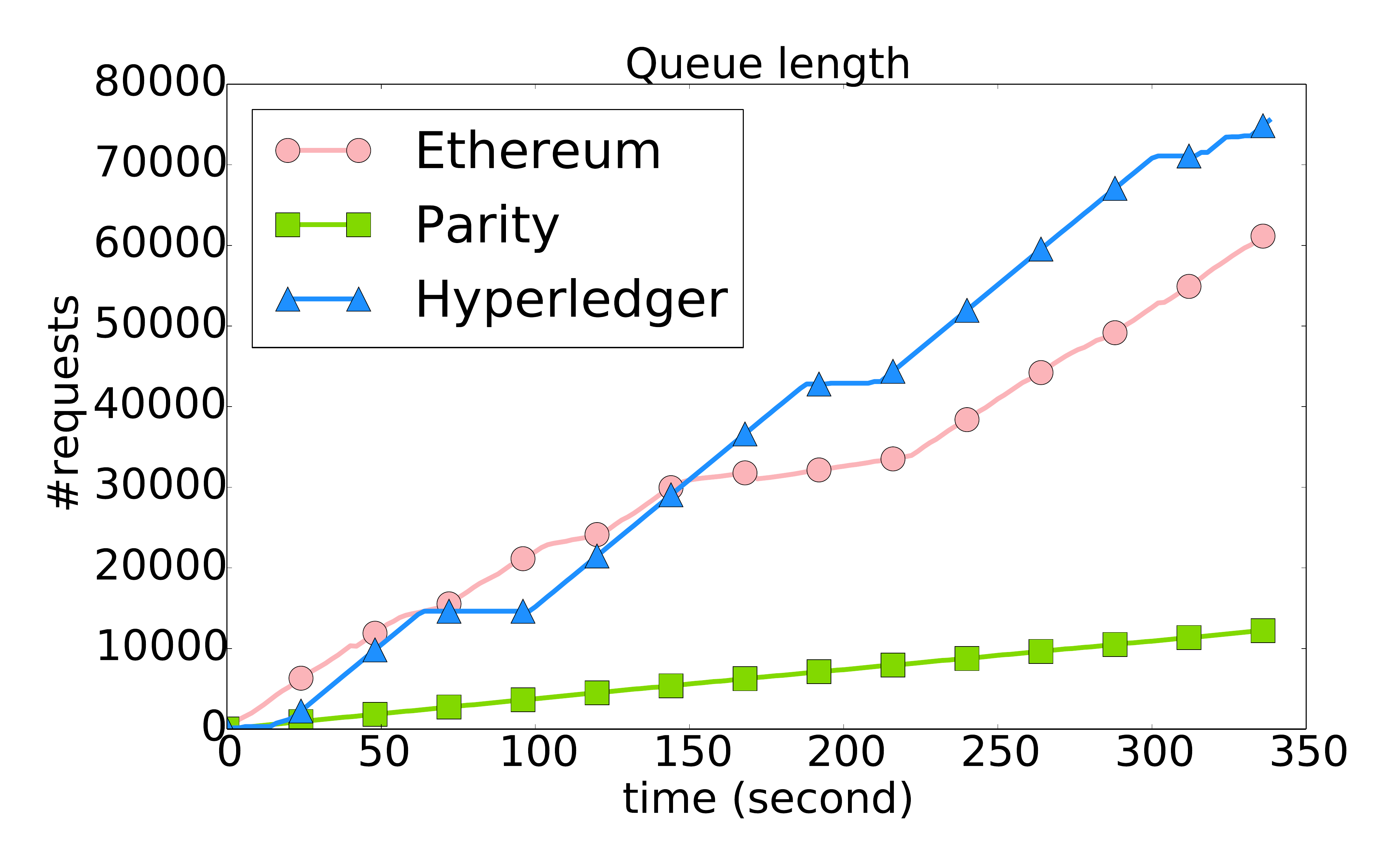}
\caption{Queue length at the client.}
\label{fig:queue_length}
\end{figure}
Figure~\ref{fig:queue_length} illustrates the request queue at the client for the settings of 20 servers and 20 clients.
The queue behavior of Ethereum reflects the normal case, i.e. the queue grew and shrank depending on how fast the
transactions are committed. Hyperledger failed to generate blocks in this case, therefore the queue never shrank.
However, there are durations in which the queue size remains constant. Furthermore, at the beginning, the queue in Hyperledger is
smaller than that in Ethereum, even though the clients are sending at the same rate. This suggests there is a bottleneck
in processing network requests at the Hyperledger servers. 

\begin{figure}
\centering
\includegraphics[width=0.42\textwidth]{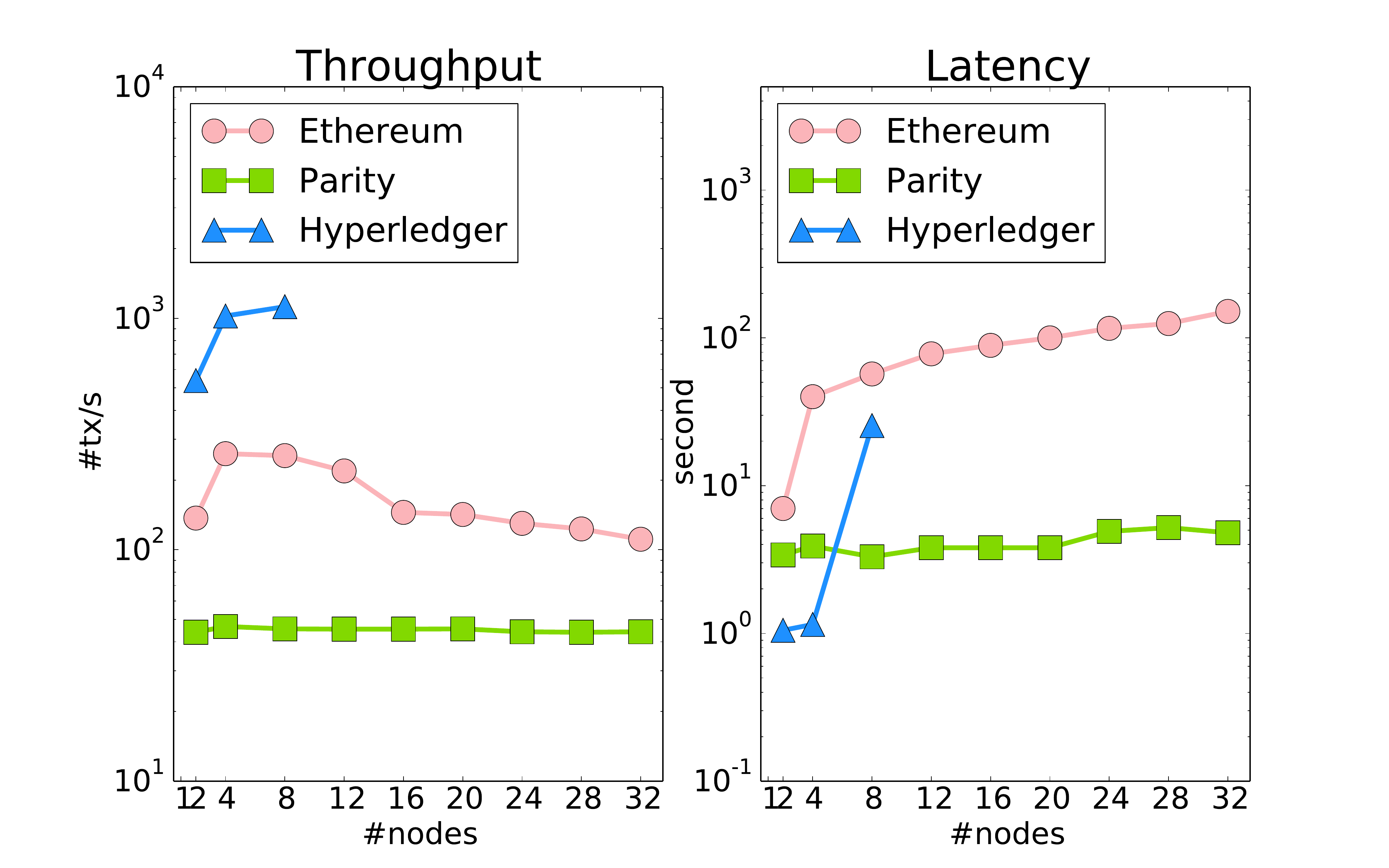}
\caption{Scalability with Smallbank benchmark.}
\label{fig:scale_sb}
\end{figure}

Figure~\ref{fig:scale_sb} illustrates the scalability of the three systems using the Smallbank benchmark. We observe similar
patterns to the YCSB benchmark (Figure~\ref{fig:scale}), except that Hyperledger failed to scale beyond 8 nodes instead
of 16.

\section{Anlaytics Smart Contract}
\begin{figure}[h]
\centering
{\footnotesize
\begin{verbatim}
type account_t struct {
    Balance     int
    CommitBlock int
}
type transaction_t {
    From string
    To   string
    Val  int
}
func Invoke_SendValue(from_account string,
        to_account string, value int) {
    var pending_list []transaction_t
    pending_list = decode(GetState("pending_list"))
    var new_txn transaction_t
    new_txn = transaction_t {
        from_account, to_account, value
    }
    pending_list = append(pending_list, new_txn)
    PutState('pending_list', encode(pending_list))
}
func Query_BlockTransactionList(block_number int)
     []transaction_t {
    return decode(GetState("block:"+block_number))
}
func Query_AccountBlockRange(account string,
        start_block int, end_block int)
     []account_t {
    version := decode(GetState(account+":latest"))
    var ret []account_t
    while true {
        var acc account_t
        acc = decode(GetState(account+":"+version))
        if acc.CommitBlock >= start_block &&
           acc.CommitBlock < end_block {
            ret = append(ret, acc)
        } else if acc.CommitBlock < start_block {
            break;
        }
        version -= 1
    }
    return ret
}
\end{verbatim}
}
\caption{Code snippet from the VersionKVStore smart contract for analytics workload (Q1 and Q2).}
\label{fig:analytic_sc} \end{figure}
Figure ~\ref{fig:analytic_sc} shows the implementation of the smart contract method that answer Q2 of the analytics
workload. To support historical data lookup, we append a counter to the key of each account. To fetch a specific version
of an account, we use key {\tt account:version}. We store the latest version of the account using key {\tt
account:latest}, and keep a {\tt CommitBlock} in the data field for every version which indicates in
which block the balance of this version is committed. To answer query that fetches a list of balance of a given  account
within a given block range, the method scans all versions of this account and returns the balance values that are
committed within the given block range. Ethereum and Parity provide JSON-PRC APIs {\tt getBalance(account, block)} to
query information of an account at a given block number.  This API fetches only one version of the account per HTTP
roundtrip, so it is less efficient than pushing the query logic to server side.

\end{document}